\documentclass[manuscript]{emulateapj}

\newcommand{\cxo}{{\sl Chandra}}
\newcommand{\ngc}{NGC~4631}

\newcommand{\xmm}{{\sl XMM-Newton}}

\begin{document}

\title{Different types of ultraluminous X-ray sources in NGC\,4631}

\author{
Roberto~Soria\altaffilmark{1}, and
Kajal~K.~Ghosh\altaffilmark{2}
}

\altaffiltext{1}{Mullard Space Science Laboratory, 
University College London, Holmbury St Mary, 
Dorking, Surrey RH5 6NT. Email: {\tt roberto.soria@mssl.ucl.ac.uk}}
\altaffiltext{2}{Universities Space Research Association,
NASA Marshall Space Flight Center, VP62, Huntsville, AL 35805, USA}

\slugcomment{Submitted to ApJ on 2008 September 2. Accepted 2009 January 27.}

\begin{abstract}
We have re-examined the most luminous X-ray sources in the 
starburst galaxy NGC\,4631, using {\it XMM-Newton}, 
{\it Chandra} and {\it ROSAT} data. 
The most interesting source is a highly variable 
supersoft ULX. We suggest that its bolometric 
luminosity $\sim$ a few $10^{39}$ erg s$^{-1}$ 
in the high/supersoft state: this is an order 
of magnitude lower than estimated in previous studies, 
thus reducing the need for extreme or exotic scenarios. 
Moreover, we find that this source was in a non-canonical 
low/soft ($kT \sim 0.1$--$0.3$ keV) state during the {\it Chandra} 
observation. By comparing the high and low state, 
we argue that the spectral properties may not be 
consistent with the expected behaviour of an 
accreting intermediate-mass black hole. 
We suggest that recurrent super-Eddington 
outbursts with photospheric expansion from a massive 
white dwarf ($M_{\rm wd} \ga 1.3 M_{\odot}$), 
powered by non-steady nuclear burning, 
may be a viable possibility, in alternative 
to the previously proposed scenario of a super-Eddington 
outflow from an accreting stellar-mass black hole.
The long-term average accretion rate required for 
nuclear burning to power such white-dwarf outbursts 
in this source and perhaps in other supersoft ULXs 
is $\approx 5$--$10 \times 10^{-6} M_{\odot}$ yr$^{-1}$: 
this is comparable to the thermal-timescale mass transfer rate 
invoked to explain the most luminous hard-spectrum ULXs 
(powered by black hole accretion).
The other four most luminous X-ray sources in NGC\,4631  
(three of which can be classified as ULXs) appear to be 
typical accreting black holes, in four different spectral 
states: high/soft, convex-spectrum, power-law with soft excess, 
and simple power-law. None of them requires 
masses $\ga 50 M_{\odot}$.

\end{abstract}


\section{Introduction}

The most luminous non-nuclear X-ray sources in nearby galaxies
occur in regions of current or recent star formation.
Some of them have X-ray luminosities exceeding the isotropic Eddintgton
luminosity $L_{\rm Edd}$ for an $\approx 10 M_{\odot}$ black hole (BH); 
they are commonly labelled ultraluminous X-ray sources (ULXs).
The conservative interpretation is that the large majority 
of ULXs are the upper end of the high-mass X-ray binary 
population, powered by an accreting BH formed from ``normal'' 
stellar processes. If so, their (apparent) extreme luminosity 
is due to any of the following three reasons, 
or to a combination of them: moderately unisotropic emission 
\citep{kin01,kin08}; mildly super-Eddington luminosities 
\citep{beg02,beg06,Ohs07}; extremely heavy stellar-mass BHs, 
with masses $\sim 30$--$70 M_{\odot}$ \citep{pak02}.
Those scenarios also require mass accretion rates 
$\dot{m} \ga 1$, where the accretion parameter 
$\dot{m} \equiv \dot{M}/\dot{M}_{\rm Edd} 
\approx \left(0.1 c^2\dot{M}\right)/L_{\rm Edd}$.
Alternatively, there is still room for the more intriguing 
hypothesis that at least some ULXs are powered 
by intermediate-mass BHs \citep{mil04}.

In the absence of direct kinematic measurements 
(because of the faintness of their optical counterparts), 
X-ray spectral and timing studies have been used 
to try and constrain BH masses in ULXs. Such model-dependent arguments 
rely on the (expected) simple scaling of characteristic variability 
timescales and disk temperatures with BH mass, and on the correspondence 
of ULX spectral states with the ``canonical'' states 
of Galactic BHs, for which the mass is accurately known. 
Unfortunately, ULXs do not appear to have the same 
state-transition behavior as Galactic BHs; for example, 
the most luminous sources are rarely found in a high/soft state, 
dominated by a standard accretion disk \citep{sor08}.
The X-ray spectra of some ULXs are dominated by an unbroken power law, 
with photon index $\Gamma \sim 1.5$--$2$, whose physical origin is still 
unclear. Others have a broad component with a steepening or 
downward curvature above $\sim 5$ keV and sometimes a small 
soft excess below $\sim 0.5$ keV; this kind of spectrum 
may come from a slim disk, or from the inner region of a standard disk, 
heavily modified by Comptonization when $\dot{m} \ga 1$.
However, there is no clear gap between the two kinds of spectra, 
and the phenomenological classification of a source 
in either class usually depends on the signal-to-noise ratio 
available in the observations.

A small subsample of ULXs stands out from this general 
spectral classification: they have a thermal spectrum with 
temperatures $\la 0.1$ keV and no emission above 1 keV. 
This is similar to the spectrum of classical supersoft sources 
in the Milky Way and Magellanic Clouds, but their luminosity 
is one or two orders of magnitude higher. 
The two most luminous supersoft ULXs are M101 ULX-1 
\citep{kon05} and NGC\,4631 X1 \citep{car07}. 
Both sources are strongly variable or transient; when in a high state, 
their blackbody luminosity is $\sim 10^{40}$ erg s$^{-1}$.
Other supersoft ULXs reaching bolometric luminosities $\ga 10^{39}$ 
erg s$^{-1}$ have been found in M81 \citep{swa02}, 
in the Antennae \citep{fab03} and (two) in NGC\,300 \citep{car06,kon03}.
Their thermal spectra can, in principle, provide tighter constraints 
on the size of the emitting region, and hence more significant 
tests for the geometry of the accretion flow 
and the nature and mass of the accretor.

While ULXs with a harder (power-law-like, slim-disk or Comptonized) 
X-ray spectrum may be interpreted as the upper end or the natural extension 
(either in BH mass or accretion rate) of stellar-mass BHs, 
supersoft ULXs appear like the upper end of nuclear-burning 
white dwarfs, which cannot be more massive 
than $\approx 1.4 M_{\odot}$. Therefore, their high apparent 
luminosities are even more difficult to explain. 
The white dwarf scenario may 
be salvaged if supersoft ULXs are seen in a transient 
outburst phase, well above the Eddington luminosity 
of a white dwarf ($\sim 10^{38}$  erg s$^{-1}$).
The other main competing scenarios for the emitting region 
in supersoft ULXs are a strong outflow from a stellar-mass 
BH accreting at a super-Eddington rate ($\dot{m} \gg 1$), 
or a standard disk around an intermediate-mass BH. 
For the disk to be so cool ($kT_{\rm in} \la 0.1$ keV), 
the BH mass needs to be $\ga 10^4 M_{\odot}$.


In this paper, we re-examine the physical interpretation 
of the supersoft ULX in NGC\,4631, by comparing the {\it XMM-Newton} 
observations taken when the source was in a more luminous state, 
with earlier {\it Chandra} observations in a lower state.
We also discuss the physical nature of the other four most luminous 
X-ray sources in the {\it XMM-Newton} dataset.

\begin{figure}[t]
\includegraphics[angle=0,width=8.5cm]{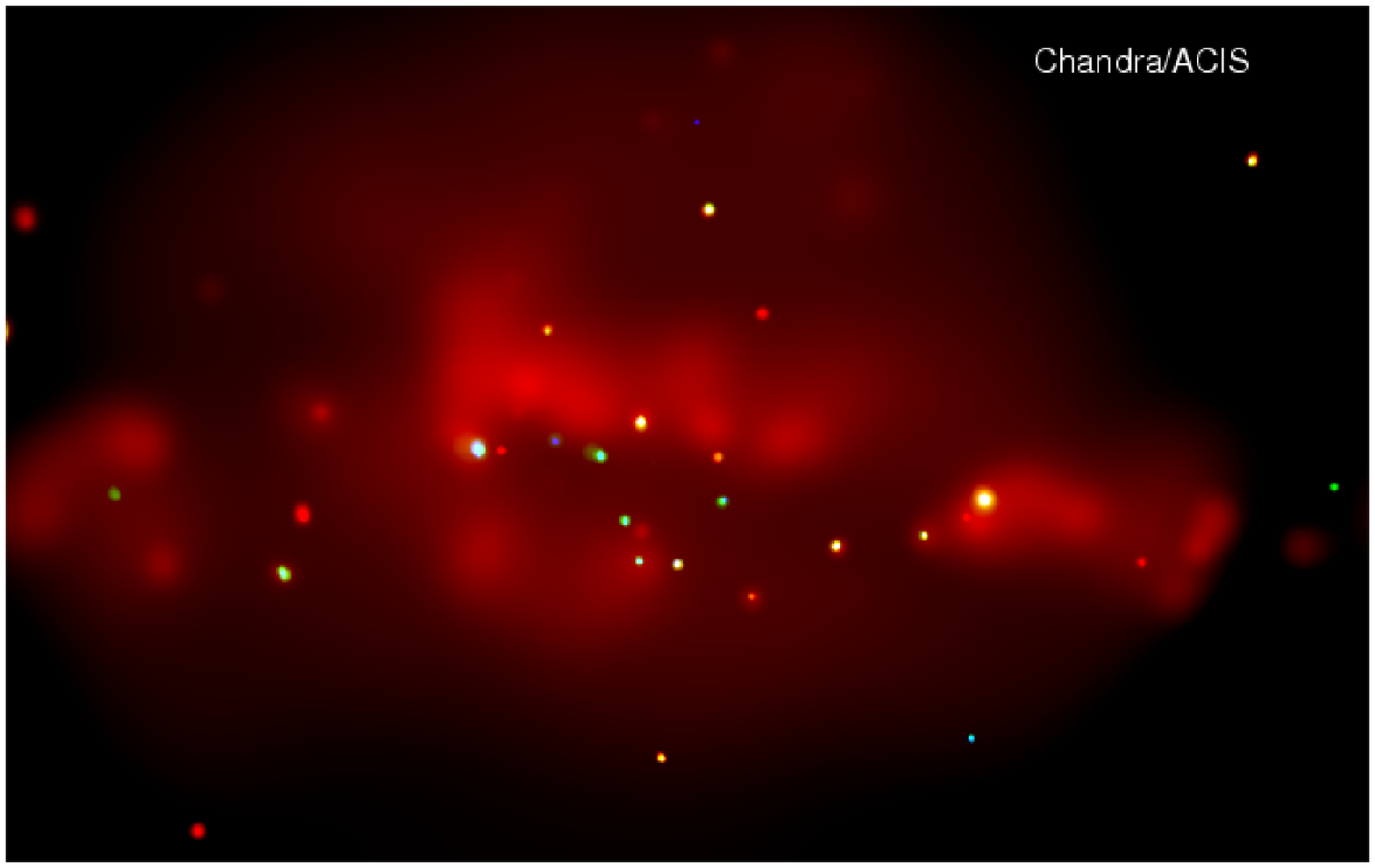}\\
\includegraphics[angle=0,width=8.5cm]{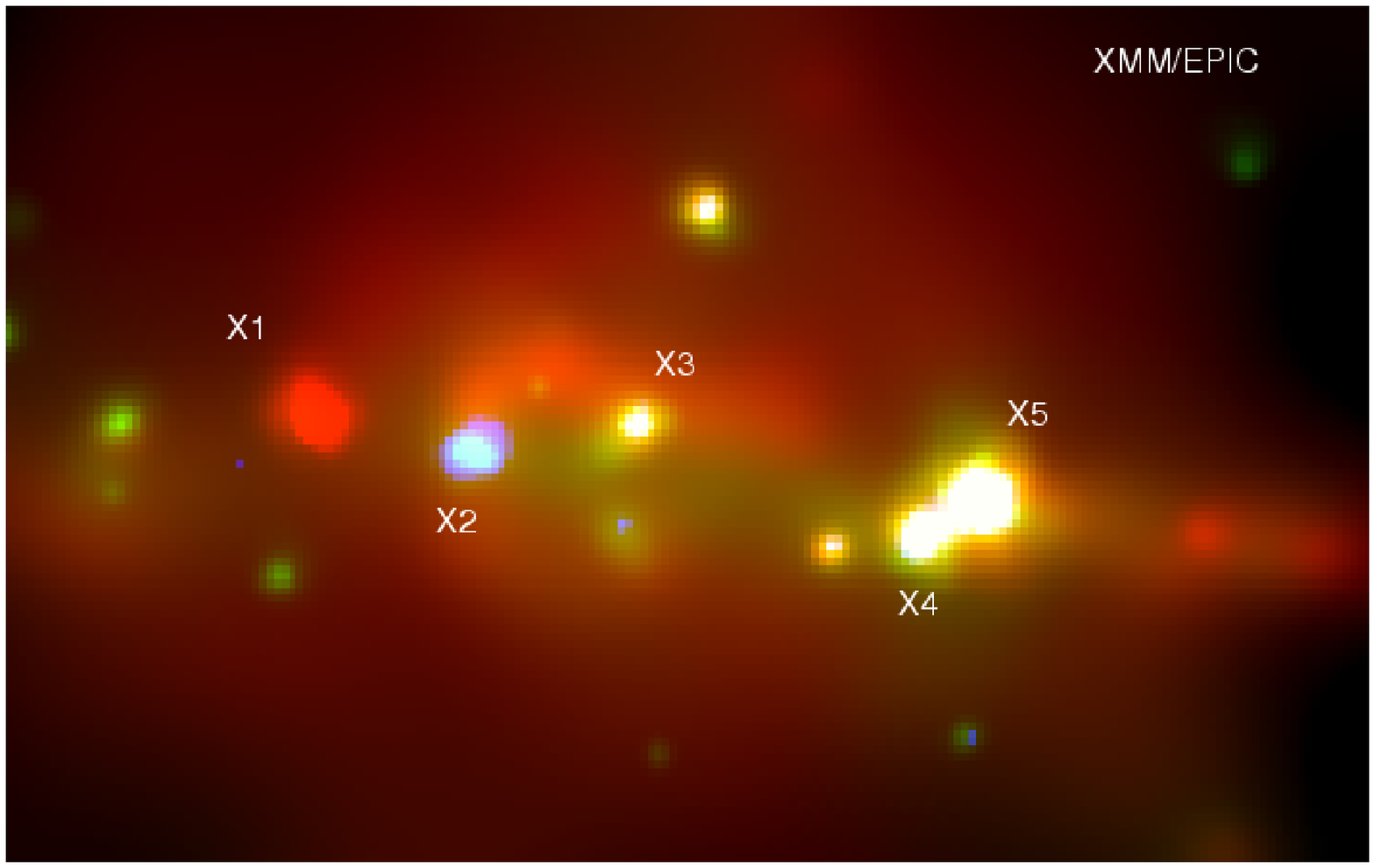}\\
\includegraphics[angle=0,width=8.5cm]{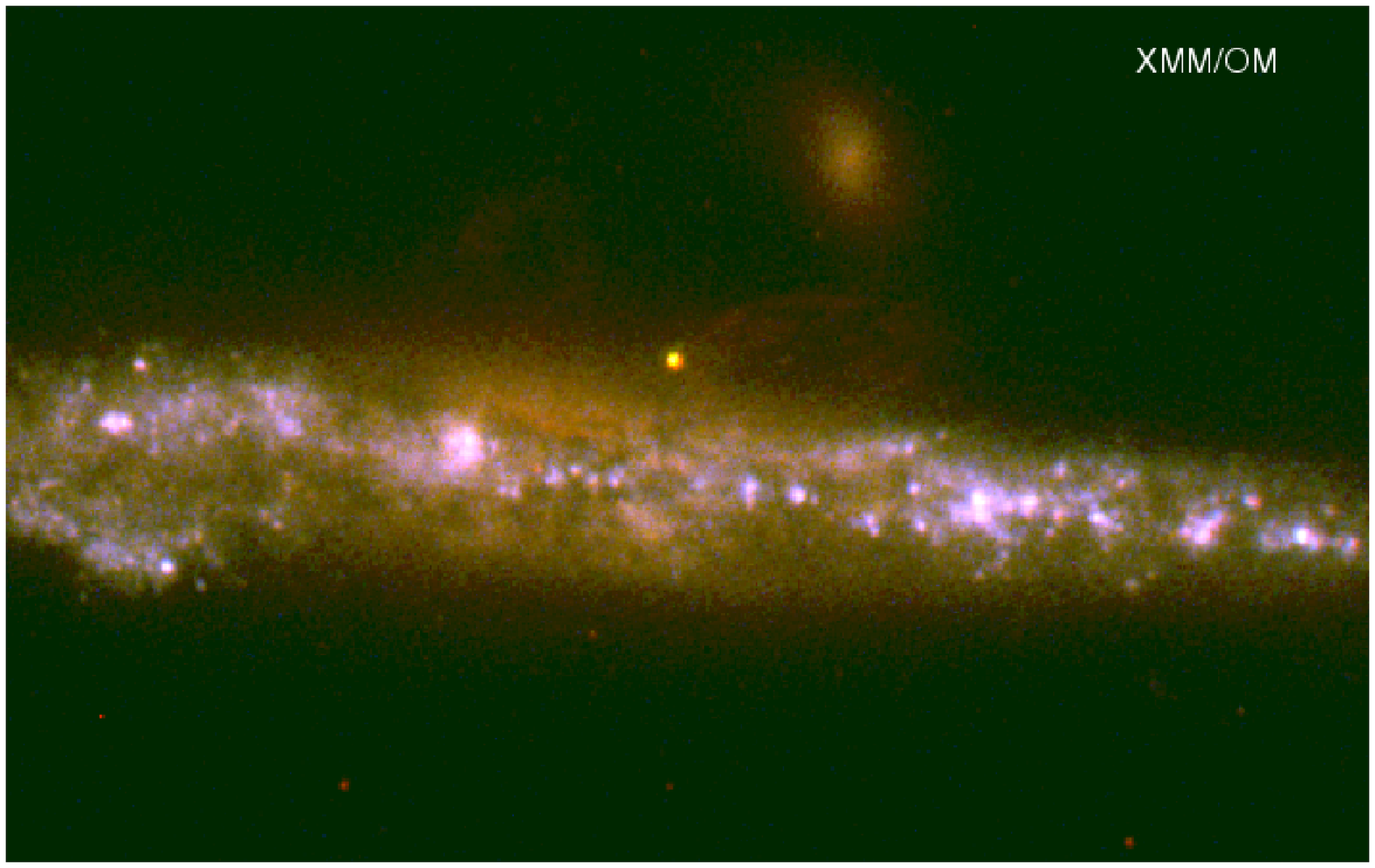}\\
\includegraphics[angle=0,width=8.5cm]{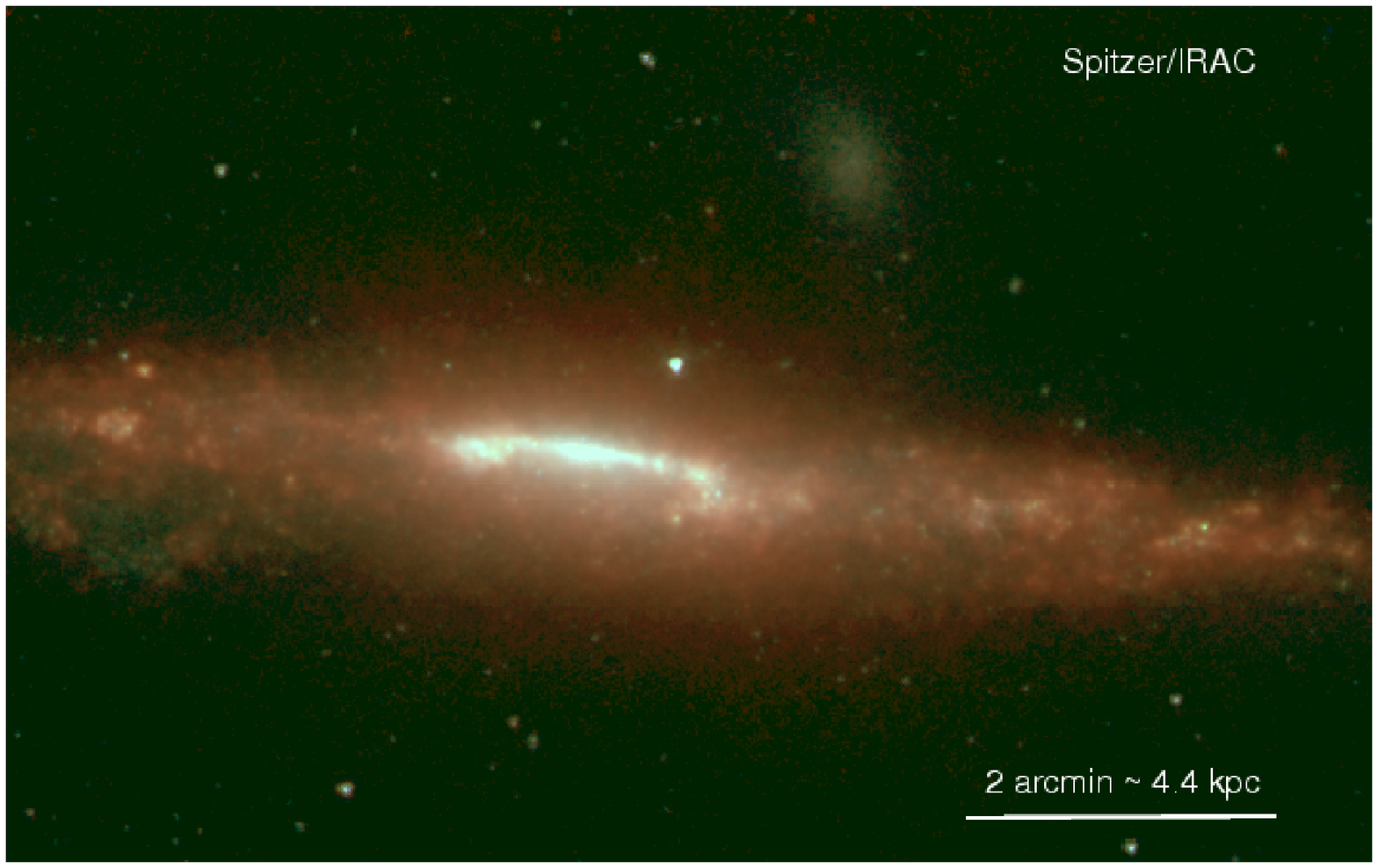}
\figcaption{Multi-band view of NGC\,4631. The true-color images 
from top to bottom are: 
{\it Chandra}/ACIS (red $= 0.3$--$1$ keV; green $=1$--$2$ keV; 
blue $=2$--$8$ keV); {\it XMM-Newton}/EPIC (red $= 0.25$--$1$ keV; 
green $=1$--$2$ keV; blue $=2$--$10$ keV), {\it XMM-Newton}/OM 
(red $= U$ filter; green $= UVW1$ filter; blue $= UVW2$ filter); 
{\it Spitzer}/IRAC (red $= 5.8$ $\mu$m; green $= 4.5$ $\mu$m; 
blue $= 3.6$ $\mu$m). The five brightest sources in 
{\it XMM-Newton}, labelled X1 through X5, are the target of 
our study. In all images, North is up and East is left.}
\label{fig:figure1}
\end{figure}


\begin{table*}
\begin{center}
\caption{Luminous X-ray sources in NGC~4631}
\begin{tabular}{lccccccr}
\hline \hline\\
\multicolumn{1}{c}{Source ID}& \it{ROSAT} ID 
     &  \multicolumn{1}{c}{R.A.} &\multicolumn{1}{c}{Dec.} 
     & \multicolumn{4}{c}{Emitted 
          Luminosity\tablenotemark{a}\tablenotemark{b} (erg s$^{-1}$)}\\[3pt]
 &&&& {\it Chandra}/ACIS & {\it XMM-Newton}/EPIC 
    & {\it ROSAT}/HRI & {\it ROSAT}/PSPC\\[3pt]
\hline\\
X1&H13&12 42 15.96 & 32 32 49.4 
         & $\sim 10^{37}$ & $3.6^{+0.5}_{-0.9} \times 10^{39}$  
         & $5.7^{+1.5}_{-1.5} \times 10^{39}$ 
         & $6.7^{+1.8}_{-1.8} \times 10^{39}$\\[3pt]
X2&-& 12 42 11.13 & 32 32 35.8 
         & $4.0^{+0.3}_{-0.3} \times 10^{39}$ 
         & $3.1^{+0.2}_{-0.2} \times 10^{39}$ &-  &- \\[3pt]
X3&-&12 42 06.07 & 32 32 46.5 
         & $3.9^{+0.2}_{-0.2} \times 10^{38}$ 
         & $4.0^{+0.2}_{-0.2} \times 10^{38}$ &-  &- \\[3pt]
X4&H8&12 41 57.35 & 32 32 03.2 
         & $3^{+1}_{-1} \times 10^{37}$ 
         & $2.1^{+0.2}_{-0.2} \times 10^{39}$ 
         & $9^{+2}_{-2} \times 10^{38}$ &$9^{+2}_{-2} \times 10^{38}$ \\[3pt]
X5&H7&12 41 55.56 & 32 32 16.9 
         & $3.8^{+0.1}_{-0.1} \times 10^{39}$ 
         & $5.0^{+0.2}_{-0.2} \times 10^{39}$ 
         & $3.5^{+0.3}_{-0.3} \times 10^{39}$ 
         & $2.9^{+0.4}_{-0.4} \times 10^{39}$ \\[3pt]
\hline
\end{tabular}
\end{center}
{\footnotesize $^a$ For the supersoft source X1:  
bolometric luminosity ($E > 13.5$ eV); for the other sources: 
unabsorbed luminosity in the $0.3$--$10$ keV band.} \\  
{\footnotesize $^b$ The error ranges listed for {\it Chandra} 
and {\it XMM-Newton} luminosities come from our spectral fitting; 
the error ranges for {\it ROSAT} luminosities include 
only the Poisson uncertainty in the HRI and PSPC count rates, 
after assuming the {\it XMM-Newton} best-fitting model 
for counts-to-flux conversion.}\\
\end{table*}

\begin{table*}
\begin{center}
\caption{Extrapolated bolometric luminosity of X1 for different fitting models}
\begin{tabular}{lccr}
\hline\hline\\
{\footnotesize XSPEC} model & $\chi^2_{\nu}$ & $kT_{\rm bb}$ (eV) 
       & $L_{\rm bol}$ (erg s$^{-1}$) \\[3pt]
\hline\\
{\tt phabs$_{\rm Gal}$*phabs*bb}  & $2.20 (87.9/40)$ 
       &  $70^{+11}_{-10}$ & $1.5\times 10^{41}$\\[3pt]
{\tt phabs$_{\rm Gal}$*phabs*(bb+gauss)}  & $1.40 (52.0/37)$ 
       &  $69^{+4}_{-4}$ & $4.5\times 10^{40}$\\[3pt]
{\tt phabs$_{\rm Gal}$*phabs*(bb+ray)}   &  $1.80 (68.3/38)$
       &  $78^{+12}_{-11}$ & $3.6\times 10^{39}$ \\[3pt]
{\tt phabs$_{\rm Gal}$*phabs*(bb+ray)*zedge}   & $1.23 (44.3/36)$ 
       &  $91^{+7}_{-9}$ & $3.6 \times 10^{39}$ \\ [3pt]{\tt phabs$_{\rm Gal}$*(ray+phabs*bb)}   &  $1.50 (57.0/38)$
       &  $66^{+14}_{-23}$& $2.0\times 10^{40}$\\[3pt]
{\tt phabs$_{\rm Gal}$*(ray+phabs*bb)*zedge}   &  $1.24 (44.6/36)$
       &  $85^{+6}_{-10}$ & $5.4\times 10^{39}$\\[3pt]
\hline\\
\end{tabular}
\end{center}
\end{table*}

\section{Observations and Data Analysis}

\ngc\ is a late-type starburst galaxy (Hubble type SB(s)d), seen nearly
edge-on (Figure 1), at a distance of $7.6$ Mpc \citep{set05}. 
In addition to a large number of giant starforming complexes,
it has one of the best examples of galactic fountains, outflows
and hot gas above the disk plane \citep{str04a,str04b,wan01}.
From its integrated far-infrared luminosity, its star-formation rate
is $\approx 3 M_{\odot}$ yr$^{-1}$ \citep{str04a,soi89,ken98}.
In the X-ray band, \ngc\  was studied with {\it Einstein} \citep{fab92},
{\it ROSAT} \citep{vog96,rea97,liu05}, {\it Chandra} \citep{wan01}
and {\it XMM-Newton} \citep{tul06a,tul06b,fen05,win06,win07,car07}.
Here, we focus on
the five brightest point-like sources in the {\it XMM-Newton} dataset 
(including four ULXs), and in particular on the supersoft ULX
\citep{car07}, whose nature is still controversial.


{\it Chandra} observations of \ngc\ with 
the Advanced CCD Imaging Spectrometer (ACIS)
were carried out on 2000 April 16, for 60 ks (ObsID 797;
Principal Investigator: Daniel Wang).
We retrieved the data from the public archive, and analysed them
with the locally-developed software tool {\footnotesize{LEXTRCT}} 
\citep{ten06}. Source detection in {\footnotesize{LEXTRCT}} 
was performed using a circular Gaussian
approximation to the point spread function (PSF), which gives higher
weight to sources with a central concentration of events. Point-source
counts and spectra were extracted from
within the 95\% encircled-energy aperture of the model PSF.
The background was extracted from annular
regions surrounding the sources, except in crowded
regions of the field where we used background regions adjacent to
the sources. The background-subtracted counts
within the source regions were scaled to obtain the aperture-corrected
count values. The background-subtracted point-source detection limit
is 14 counts for the 2.8 minimum sigal-to-noise ratio (S/N)
threshold and a minimum 5 $\sigma$ above background.
For timing analysis, we binned the X-ray light curves of the brightest
sources into 1000-s bins, and computed $\chi^{2}$ tests against
a constant flux hypothesis.
For spectral analysis, we generated spectral redistribution matrices
and ancillary response files with the \cxo\ X-ray Center
software {\footnotesize CIAO} version 3.4.
We then used {\footnotesize XSPEC} version 12.0 \citep{arn96} 
to fit the point-source spectra.

A 55-ks \xmm\ observation with the European Photon Imaging Camera 
(EPIC) was carried out on 2002 June 28
(ObsID 0110900201; Principal Investigator: Michael Watson).
We downloaded the public-archive data and processed them with
the \xmm\ Science Analysis System ({\footnotesize{SAS}}) 
version 6.5.0. We used
{\footnotesize LEXTRCT} for source detection, and standard 
{\footnotesize{XMMSELECT}} tasks within the {\footnotesize{SAS}} 
for source and background region extraction. The radius of the source 
extraction regions was $20\arcsec$, except for X4, where 
we used a $15\arcsec$ radius to reduce contaminations from 
the nearby brighter source X5. Background extraction 
regions were chosen around the source regions, in a suitable 
way to avoid contamination. After building 
response and ancillary response files with 
{\footnotesize{rmfgen}} and {\footnotesize{arfgen}}
we used {\footnotesize XSPEC} for spectral analysis
of the brightest sources (Table 1).
To improve the signal-to-noise ratio, we co-added 
the EPIC pn and MOS spectra,
with suitably averaged response functions,
using the method of \citet{pag03}.

To investigate the long-term variability of the ULXs,
we also re-analyzed the archival {\it ROSAT}/HRI and PSPC observations
carried out between 1991 December and 1992 December \citep{vog96,rea97}.
We applied astrometric corrections to the {\it ROSAT} data using the 
{\it Chandra}
source positions. We extracted source counts from circular regions
of 30\arcsec radius, and background counts from source-free circular
regions of 3\arcmin radius.
We used {\footnotesize WebPIMMS} with the best-fitting spectral 
parameters from the {\it XMM-Newton} study 
to convert {\it ROSAT} source count rates
into fluxes. Three (X1=H13, X4=H8, and X5=H7; see 
Table 1 and \citet{vog96}) 
of the five luminous targets of this study were also found in {\it ROSAT}, 
with some variability over the various exposures.



\begin{table}
\begin{center}
\caption{Best-fit parameters for the coadded EPIC pn and MOS 
spectrum of X1 in its high/supersoft state. 
Spectral model: {\tt phabs*phabs*(raymond-smith + bbody)*zedge}. 
Values in brackets were fixed. Errors are 90\%
confidence levels for 1 interesting parameter ($\Delta \chi^2 =
2.7$). }
\vspace{0.2cm}
\begin{tabular}{lr}
\tableline\tableline\\
Parameter & {\it XMM-Newton} Value \\
\tableline\\
$N_{\rm {H,Gal}}$\tablenotemark{a} & $(1.3 \times 10^{20})$\\[6pt]
$N_{\rm {H}}$ & $2.4^{+0.3}_{-0.3} \times 10^{21}$\\[6pt]
$kT_{\rm{rs}}$ (keV) & $0.44^{+0.06}_{-0.06}$ \\[6pt]
$Z(Z_{\odot})$ & $(1.0)$ \\[6pt]
$K_{\rm{rs}}$\tablenotemark{b} & $9.2^{+1.9}_{-2.0} \times 10^{-6}$\\[6pt]
$kT_{\rm{bb}}$ (keV) & $0.091^{+0.07}_{-0.09}$ \\[6pt]
$K_{\rm{bb}}$\tablenotemark{c} & $5.7^{+0.5}_{-0.5}  \times 10^{-6}$\\[6pt]
$E_{\rm{edge}}$ (keV) & $1.01^{+0.01}_{-0.07}$ \\[6pt]
$\tau_{\rm{edge}}$  & $2.0^{+1.2}_{-0.7}$ \\[6pt]
\tableline\\
$\chi^2$/dof & $1.23 (44.3/36)$ \\[3pt]
\tableline\\
$f_{\rm 0.3-10}$\tablenotemark{d} &$3.3^{+0.1}_{-0.1} \times 10^{-14}$\\[6pt]
$f_{\rm 0.3-10, rs}$\tablenotemark{e} &$8.4^{+2.1}_{-1.8} \times 10^{-15}$\\[6pt]
$f_{\rm 1-10}$\tablenotemark{f} &$1.4^{+0.1}_{-0.1} \times 10^{-15}$\\[6pt]
$f_{\rm 1-10, rs}$\tablenotemark{g} &$1.0^{+0.2}_{-0.2} \times 10^{-15}$\\[6pt]
$L_{\rm bol}$\tablenotemark{h} & $3.6^{+0.9}_{-0.5} \times 10^{39}$\\[6pt]
$L_{\rm bol, rs}$\tablenotemark{i} & $2.6^{+0.7}_{-0.5} \times 10^{38}$\\[6pt]
\tableline\\
\end{tabular}
\vspace{-0.2cm}
\tablenotetext{a}{From \citet{K05}. Units of cm$^{-2}$.}
\tablenotetext{b}{Raymond-Smith model normalization $K_{\rm rs} =
10^{-14}/\{4\pi\,[d_A\,(1+z)]^2\}\,
\int n_e n_H {\rm{d}}V$, where $d_A$ is
          the angular size distance to the source (cm), $n_e$ is the electron
          density (cm$^{-3}$), and $n_H$ is the hydrogen density (cm$^{-3}$).}
\tablenotetext{c}{Blackbody model normalization $K_{\rm{bb}} = L_{39}/D^2_{10}$ 
          where $L_{39}$ is the source luminosity in units of $10^{39}$ erg s$^{-1}$ 
	  and $D_{10}$ is the distance in units of $10$ kpc.}
\tablenotetext{d}{Observed flux in the $0.3$--$10$ keV band; 
          units of erg cm$^{-2}$ s$^{-1}$.}
\tablenotetext{e}{Observed flux in the $0.3$--$10$ keV band, in the Raymond-Smith 
          component; units of erg cm$^{-2}$ s$^{-1}$.}
\tablenotetext{f}{Observed flux in the $1$--$10$ keV band; 
          units of erg cm$^{-2}$ s$^{-1}$.}
\tablenotetext{g}{Observed flux in the $1$--$10$ keV band, in the Raymond-Smith 
          component; units of erg cm$^{-2}$ s$^{-1}$.}
\tablenotetext{h}{Unabsorbed luminosity for $E>13.6$ eV; 
          units of erg s$^{-1}$.}
\tablenotetext{i}{Unabsorbed luminosity for $E>13.6$ eV, in the Raymond-Smith 
          component; units of erg s$^{-1}$.}\\
\end{center}
\end{table}

\section{The supersoft ULX}

X1 was detected as a luminous supersoft source 
during the {\it ROSAT}/PSPC observations of 1991 December 15 -- 
1992 January 04 (a total of 18.4 ks; see Fig.~4 and Table 1 
in \citet{vog96}). It was not detected in the shorter (3.5 ks) 
{\it ROSAT}/PSPC observations of 1992 May, suggesting a count-rate 
decline by at least a factor of 3. It was detected again by 
{\it ROSAT}/HRI in 1992 December; however, since the HRI does not 
provide spectral information, we cannot tell whether it was again 
in a super-soft state.
In the {\it Chandra} observation from 2000 April 16, the source 
was faint and soft. Finally, in the {\it XMM-Newton} observations 
of 2002 June 28, X1 appeared again as a luminous supersoft source 
\citep{car07}, in a similar state to the 1991 detection.
Here, we briefly summarize the spectral results when the source 
was in a high state, and then compare them with the low-state 
observation.

\subsection{High state}
We re-extracted the {\it XMM-Newton} data and coadded 
the pn and MOS spectra with a suitably averaged 
response function. This is equivalent to fitting them simultaneosly, 
but provides a better signal-to-noise ratio for discrete features.
We recover the result of \citet{car07}, with 
a spectrum dominated by a soft blackbody component, 
plus residual features (both in emission and absorption) 
especially at $\approx 0.7$--$1.2$ keV (Figure 2). 
Such features are already evident in each individual 
EPIC spectrum, as plotted in Figure 4 of \citet{car07}, 
and become more significant when the spectra from all three 
detectors are combined. The visual impression of such 
systematic residuals is confirmed by the fit statistics:
the best-fitting absorbed blackbody model has 
$\chi^2 = 2.20(87.9/40)$ and can be safely rejected.

The residual emission and absorption features 
may not seem to affect the bolometric luminosity 
significantly, compared with the dominant blackbody 
emission. However, different models for such 
residual components 
have the effect of shifting the fitted temperature 
of the blackbody component between $\approx 65$ and $90$ eV, 
with a dramatic effect on the extrapolated, unabsorbed 
bolometric luminosity. Some examples are summarized in Table 2. 
When we model the deviations from a pure 
blackbody spectrum with only emission components 
(e.g., a Gaussian as in \citet{car07}, 
or an optically-thin thermal plasma), we find an extrapolated 
luminosity $\approx 2$--$5 \times 10^{40}$ erg s$^{-1}$.
However, we also find that there is a statistically-significant 
absorption feature at $E = 1.0 \pm 0.1$ keV. This may be 
analogous to the Fe-L absorption edges found in some 
Seyfert galaxies \citep{bol03}.
Ignoring this edge 
leads to apparently lower blackbody temperatures 
and therefore higher extrapolated bolometric luminosities.
When we include this edge in our models 
(Tables 2 and 3), we obtain bolometric luminosities 
as low as $\approx 4$--$5 \times 10^{39}$ erg s$^{-1}$, 
depending on whether we assign the same, high 
intrinsic absorption to both the optically-thin 
and optically-thick thermal components, or 
only to the latter. On the other hand, the isotropic 
emitted luminosity $2 \times 10^{39}$ erg s$^{-1}$ 
in the $0.3$--$10$ keV band provides a solid lower limit 
to the bolometric luminosity.

In our best-fitting model (Table 3), the optically-thin emission 
component contributes less than 1/10 of the extrapolated 
bolometric luminosity, but about 1/4 of the observed flux 
in the full $0.3$--$10$ keV band,  
about half of the unabsorbed luminosity at energies $> 0.7$ keV, 
and 2/3 of the unabsorbed luminosity at energies $>1$ keV.
This explains why modelling such component has a great effect 
on the fit parameters and inferred luminosity.
The unabsorbed luminosity of the optically-thin thermal plasma 
is $\approx 2.0 \times 10^{38}$ erg s$^{-1}$ 
(bolometric luminosity $\approx 2.6 \times 10^{38}$ erg s$^{-1}$).

The 1991 {\it ROSAT}/PSPC spectrum is also 
dominated by a blackbody component at 
$kT_{\rm bb} \la 0.1$ keV \citep{rea97}, 
and a {\footnotesize WebPIMMS} estimate suggests 
it may have similar luminosity 
to the 2002 {\it XMM-Newton} spectrum. However, the {\it ROSAT} 
data do not have enough counts and spectral resolution 
to constrain the temperature (and therefore the extrapolated 
emitted luminosity) more accurately. If we assume that 
the spectral model was the same as in the  
{\it XMM-Newton} observation (choosing 
for example the model listed in Table 3), and leave 
only the relative normalization free between 
the two epochs, we find that the two spectra 
are indeed consistent with being very similar (Figure 3),  
with a {\it ROSAT} flux normalization $1.33^{+0.33}_{-0.30}$ times 
higher than for {\it XMM-Newton}.

The best-fit blackbody temperature ($kT_{\rm bb} = 0.09 \pm 0.01$ keV) 
and (extrapolated) blackbody luminosity 
($L_{\rm bb} \approx 4 \times 10^{39}$ erg s$^{-1}$ in 2002, 
and $ \approx 5 \times 10^{39}$ erg s$^{-1}$ in 1991) 
correspond to a characteristic radius $\approx 2.1 \times 10^9$ 
cm or $\approx 2.4 \times 10^9$ cm (in 2002 and 1991, 
respectively) for the optically-thick emitting surface.
We should also keep in mind that a simple blackbody approximation 
tends to overestimate the true luminosity of supersoft sources 
by a factor of a few \citep{kah97}, 
so the true luminosity may be even lower.

\begin{figure}
\includegraphics[angle=-90,width=1\columnwidth]{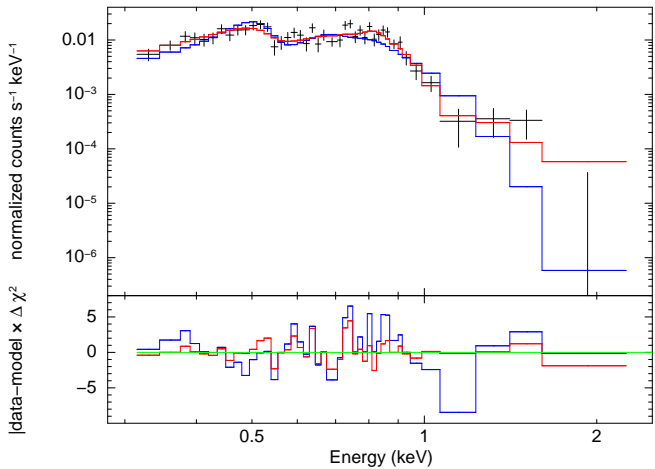}
\figcaption{Model fits and $\chi^2$ residuals 
for the combined {\it XMM-Newton}/EPIC 
spectrum of X1: simple blackbody (blue curve; $\chi^2_{\nu} \approx 2.2$), 
and blackbody plus thermal-plasma emission plus absorption edge 
(red curves; $\chi^2_{\nu} \approx 1.2$; best-fitting parameters 
in Table 3).
}
\label{fig:figure2}
\vspace{0.3cm}
\end{figure}

\subsection{Low state}
From the {\it Chandra} data in the low state, we found $\approx 13.5$ net 
ACIS counts, corresponding to a count rate of $(2.3 \pm 0.7) 
\times 10^{-4}$ ct s$^{-1}$ (a factor of two less 
than estimated by \citet{car07}). The breakdown 
of the net counts in different energy bands is:  
$\approx 12.2$ counts at $0.3$--$1.5$ keV, and 
$\approx 1.3$ counts at $1.5$--$8.0$ keV.
Dividing the interval into three bands, we get:  
$\approx 10$ counts at $0.3$--$1.0$ keV,  
$\approx 2.5$ counts at $1.0$--$2.0$ keV,  
$\la 1$ counts at $2.0$--$8.0$ keV.

So, although there are not enough counts for 
detailed spectral fitting, we have at least a strong 
indication that the source was very soft 
(or ``quasi-soft'' in the definition of \citet{dis04}). 
Even if we assume no intrinsic absorption, the count 
distribution rules out a power-law spectrum with 
photon index $\Gamma \la 2$ at the 90 per cent confidence level 
(using the Cash statistic, \citet{cas79}). 
The optically-thin thermal plasma component 
fitted to the high-state spectrum is clearly 
inconsistent with the lower flux detected 
in the low state (Figure 4). 
More generally, optically-thin thermal plasma models 
(at fixed solar metallicity) 
are also ruled out at the 90 per cent confidence level 
(the best-fitting model has a Cash-statistic parameter $= 21.8/14$)
Soft, optically-thick thermal emission (blackbody or 
disk-blackbody) is a much better model 
for the observed count distribution.
We estimate a disk-blackbody 
temperature $kT_{\rm in} = 0.22^{+0.32}_{-0.13}$ keV 
(90 per cent confidence limits; 
Cash-statistic fit parameter $= 13.6/14$)
or a simple blackbody temperature 
$kT_{\rm bb} = 0.18^{+0.13}_{-0.10}$ keV 
(Cash-statistic fit parameter $= 13.2/14$).
The best-fitting unabsorbed luminosity 
is $\approx 10^{37}$ erg s$^{-1}$, although 
the 68 per cent confidence interval includes 
values as high as $\approx 10^{38}$ erg s$^{-1}$; 
but in any case, the luminosity is lower than in the supersoft state, 
as intuitively expected from a simple scaling of the count rate 
between the {\it Chandra} and {\it XMM-Newton} observations.
From the few detected counts, we cannot strongly rule out 
the alternative possibility that the lower count rate 
in the {\it Chandra} observation is also or mostly due 
to a much higher intrinsic absorption 
($N_{\rm H} \ga 10^{22}$ cm$^{-2}$) 
of the same, luminous super-soft component. 
But in the absence of independent evidence for that dramatic 
change in absorption, and by analogy with the behaviour 
of other accreting systems (including the supersoft source 
in M101 mentioned earlier, \citet{kon05}), 
here we consider the low-state scenario 
as the most plausible.

The absence of the optically-thin thermal plasma component 
in low state is somewhat puzzling. The rapid change between 2000 and 2002 
suggests that it was not due to diffuse hot gas at large distances 
from the source, but was instead directly associated with the high state 
or outburst. The PSF and source extraction region are of course 
much larger for {\it XMM-Newton}, which may suggest a contamination 
from Galactic-scale diffuse emission; however, we tested this possibility 
using the {\it Chandra} images, and we do not find evidence of local 
enhancements in the diffuse soft emission. Different choices 
of background extraction regions in {\it XMM-Newton} do not remove 
this component. In conclusion, we suggest that the most likely 
explanation at this stage is that the line emission and absorption edge 
are really associated with the compact source.
If the emission is due to an expanding BH wind or white dwarf 
photosphere (scenarios outlined in Sections 4.2 and 4.3)
we speculate that this may be evidence of an optically thick 
and optically thin component in the outflow.

\begin{figure}
\includegraphics[angle=-90,width=1\columnwidth]{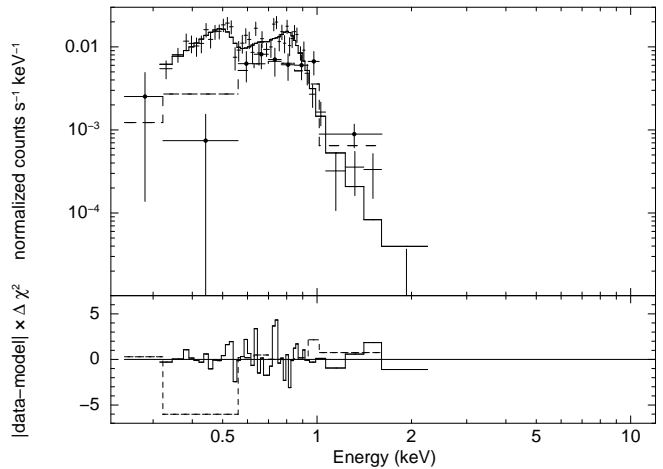}
\figcaption{Combined fit to the {\it XMM-Newton}/EPIC 
(solid line) and {\it ROSAT}/PSPC (dashed line)  
binned spectra for the supersoft source X1 in its high state 
(datapoints and $\chi^2$ residuals). 
The {\it XMM-Newton} data are from 2002 June; the {\it ROSAT} 
data from 1991 December -- 1992 January. The main component 
of the best-fitting 
model (Table 3) is an absorbed blackbody with 
$kT_{\rm bb} = 0.08 \pm 0.02$ keV; the normalization of the 
{\it ROSAT} spectrum is a factor $\approx 1.3$ higher 
than that of the {\it XMM-Newton} spectrum. 
}
\label{fig:figure3}
\vspace{0.3cm}
\end{figure}


\begin{figure}
\includegraphics[angle=-90,width=1\columnwidth]{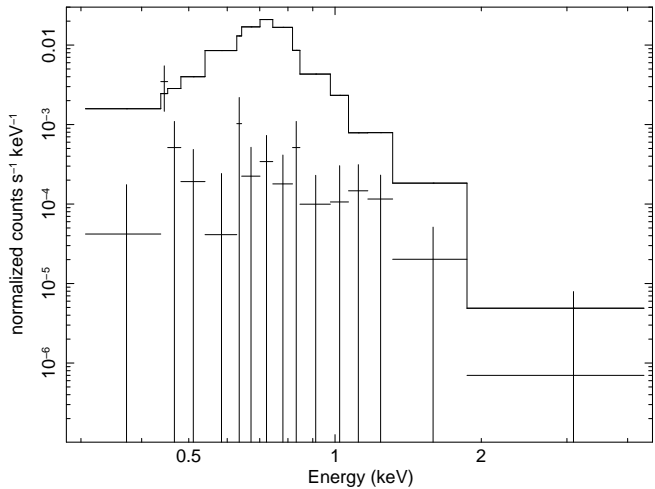}
\figcaption{Datapoints from the {\it Chandra} observation 
in the low state (each bin corresponding to a single detected count) 
compared with the expected contribution from the optically-thin 
thermal component alone, using the best-fitting parameters 
of the high state (Table 3).
}
\label{fig:figure4}
\vspace{0.3cm}
\end{figure}

\section{Physical interpretation of X1}
Three scenarios have been considered for this source 
(and for the handful of similar ones in other nearby galaxies), 
based on the properties of its high/supersoft state 
\citep{car07}: standard disk emission from 
an intermediate-mass BH; super-Eddington 
outflows from a stellar-mass BH; nuclear-burning white dwarf.
We discuss here how the low-state appearance provides 
stronger constraints.

\subsection{Intermediate-mass BH}

In the standard disk model, 
the BH mass can be expressed as a function of inner-disk 
temperature and bolometric disk luminosity (or {\footnotesize {diskbb}} 
normalization constant): $M \sim T_{\rm in}^{-2} L_{\rm disk}^{1/2}$ 
\citep{mak00}. 
Assuming $R_{\rm in} = 6M$ and imposing 
$N_{\rm H} \ge 1.3 \times 10^{20}$ cm$^{-2}$ 
(Galactic line of sight), we used the Cash-statistics fit 
to the {\it Chandra} data, and we derived a contour plot 
of the acceptable region in the 
BH-mass versus disk-luminosity space (Figure 5). We do not 
have enough net counts to determine the 90\% confidence 
contours, but we can at least identify a ``most plausible'' 
(68\% confidence contours) region in that plane. 
The large uncertainty in the inner-disk temperature 
(from $\sim 0.1$ to $\sim 0.5$ keV) is reflected 
in a large range of masses and unabsorbed luminosities.
Regardless of BH mass, the luminosity 
appears to be always $< 0.01 L_{\rm Edd}$.
However, there are observational 
and theoretical arguments against a disk-dominated 
low/soft state at $L < 0.01 L_{\rm Edd}$. 
At those luminosities, accreting BHs are generally
found in a power-law-dominated low/hard state, 
which is ruled out in this case.
Hence, we suggest that the low/soft appearance 
of the source in the {\it Chandra} observation 
does not favour the disk-blackbody model.

Moreover, the fitted inner-disk temperature in the low/soft 
state appears to be similar or slightly higher 
than the thermal temperature in the high/supersoft 
state (\citet{car07}, and Table 2). 
If the thermal emission is due to an accretion disk, 
we expect the disk to be cooler, when 
the net count rate (a proxy for the accretion rate and 
emitted luminosity) is two or three orders of magnitude lower; 
$T_{\rm in} \sim L_{\rm disk}^{1/4}$ 
in a standard disk. A transition 
between high/supersoft and low/soft states is inconsistent 
with the well-studied behaviour of BH accretion disks.
In conclusion, the comparison of low- and high-state 
spectral data does not favour the intermediate-mass 
BH scenario.

\subsection{Stellar-mass BH}

The high/supersoft state is consistent with 
thermal emission scattered and collimated by 
a massive, optically-thick disk outflow, 
launched at the spherization 
radius around a stellar-mass BH, when the accretion rate 
exceeds the Eddington limit \citep{sha73,pou07,kin08}.
We have already noted that in the {\it Chandra} observation, 
the emitted luminosity may be two or three orders of magnitude 
lower. Hence, we do not expect the super-Eddington 
outflow to be present in that lower state. 

In the canonical scheme of BH accretion states, inferred 
mostly from the study of stellar-mass Galactic 
BH binaries, moderately active (sub-Eddington) BHs are either 
in the high/soft state (dominated by a disk-blackbody component with 
$kT_{\rm in} \approx 1$ keV), or in the low/hard state 
(power-law component with $\Gamma \sim 1.5$--$2$).
Neither state is consistent with the observed 
{\it Chandra} spectrum. And conversely, no stellar-mass BHs 
have been observed in a disk-dominated 
state with $kT_{\rm in} \sim 0.2$ keV 
and $L_{\rm X} \sim 10^{37}$--$10^{38}$ erg s$^{-1}$.
If the supersoft component in the 
higher state is attributed to a disk outflow during 
super-critical accretion, we would have to conclude that
the lower-state spectrum does not look like 
a canonical state for stellar-mass BHs.

An alternative possibility we should consider is that 
the sequence of high and low states is not due to accretion 
state transitions, but to a long-term precession of the binary 
system, such that the moderatley collimated outflow moves 
in and out of our line of sight. When we are looking down the outflow, 
we may be observing the undisturbed, cooler (soft spectrum) 
outer disk, at radii larger than the spherization radius.
However, a difficulty of this interpretation is that 
the predicted isotropic luminosity of the outer standard disk 
down to the spherization radius should be 
$\approx L_{\rm Edd} \approx 10^{39}$ erg s$^{-1}$ 
for a stellar-mass BH. This is much higher than observed.

In conclusion, we suggest that the stellar-mass BH scenario, 
although still viable, has not yet provided a perfectly self-consistent 
interpretation for this source, or at least requires a new kind 
of accretion-state behaviour or accretion-disk structure, 
so far unobserved in stellar-mass BHs.

\begin{figure}
\includegraphics[angle=-90,width=1\columnwidth]{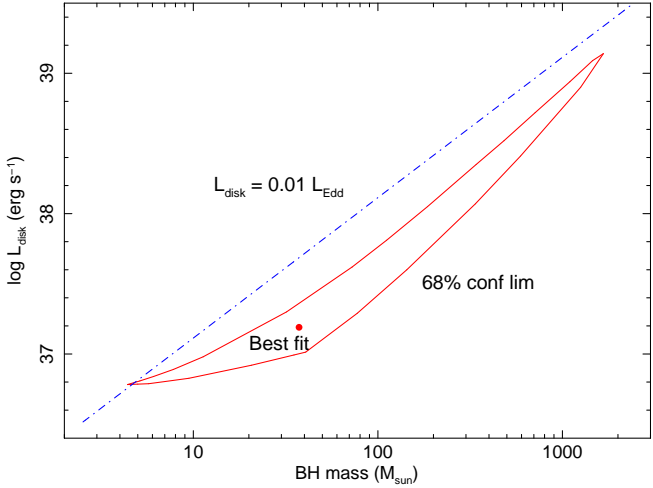}
\figcaption{Contour plot for the 68\% confidence region 
in the BH mass versus bolometric disk luminosity plane, 
for X1 in its low/soft state ({\it Chandra} 
observation), if its spectrum is fitted 
with an absorbed disk-blackbody. The dash-dotted line 
corresponds to $L_{\rm disk} = 0.01 L_{\rm Edd}$, 
a conventional threshold below which the disk 
is not expected to be the dominant emitting component 
in accreting BHs (and therefore a disk-blackbody 
model would not be self-consistent).}
\label{fig:figure5}
\vspace{0.3cm}
\end{figure}


\begin{figure}
\includegraphics[angle=-90,width=1\columnwidth]{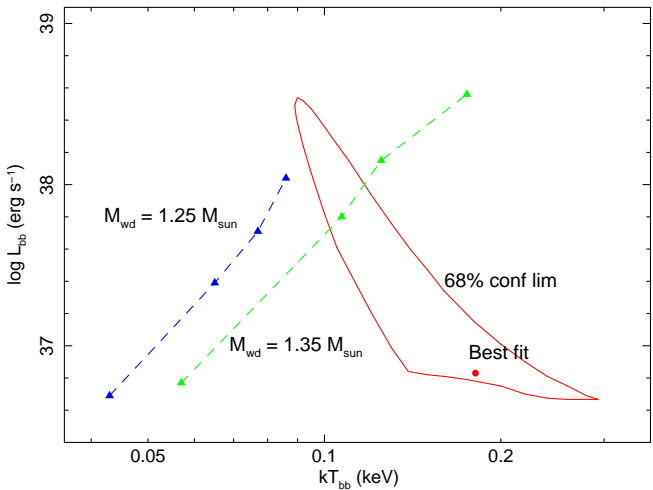}
\figcaption{Contour plot for the 68\% confidence region 
in the BH mass versus bolometric luminosity plane, 
for X1 in its low/soft state ({\it Chandra} 
observation), if its spectrum is fitted 
with an absorbed blackbody. The dashed lines correspond 
to the predicted effective temperatures and luminosities 
for massive white dwarfs ($1.25 M_{\odot}$ and $1.35 M_{\odot}$) 
during phases of surface hydrogen burning, 
for different values of mass accretion rates, 
increasing from $1.6 \times 10^{-8} M_{\odot}$ yr$^{-1}$ 
to $8.0 \times 10^{-7} M_{\odot}$ yr$^{-1}$ 
(see \citet{sta04} for detailed explanations of   
those sequences).}
\label{fig:figure6}
\vspace{0.3cm}
\end{figure}


\subsection{Nuclear-burning white dwarf}

The observed temperature and luminosity of X1 
in the low/soft state is consistent  
with the thermal emission from surface hydrogen burning 
on a massive white dwarf, $M_{\rm wd} \approx 1.3$--$1.35$  
(Figure 6).
Such process can occur at accretion rates just below 
the steady burning rate ($\sim$ a few $10^{-7} M_{\odot}$ 
yr$^{-1}$: \citet{sta04,kah04}. The white 
dwarf is sufficiently hot that hydrogen burns on its surface 
immediately as it is accreted; for this reason, 
the system does not go through classical-nova outbursts.
A blackbody luminosity $\approx 10^{38}$ erg s$^{-1}$ 
at a temperature $kT_{\rm bb} \sim 0.1$ keV 
are well within the 68\% confidence limit of the 
{\it Chandra} detection. These values correspond 
to a radius $\sim 3000$ km for a spherical emitter, 
consistent with the radius of a $1.35$-$M_{\odot}$ 
white dwarf. 

However, the high/supersoft phases cannot be explained 
with steady surface burning. It was found from numerical simulations 
\citep{sta04} that 
steady surface hydrogen burning on a 
$1.35$-$M_{\odot}$ white dwarf can occur only
for accretion rates $\la 10^{-6} M_{\odot}$ yr$^{-1}$; 
above that limit, the photosphere of the white 
dwarf expands to large radii and shuts off 
accretion. This may be what is happening in X1, 
with a sequence of outbursts in between phases 
of more steady surface burning (at reduced accretion rate) 
or simply of surface cooling.
The estimated radii (a factor of 10 higher 
than in the low/soft state, Section 3.1) 
and blackbody temperatures (a factor of two lower)
fitted to the high/supersoft spectrum 
are consistent with photospheric expansion.

From the sequence of X-ray observations 
between 1991 and 2002, we speculate that 
X1 is in a transient supersoft state about  
half of the time. Since the emitted luminosity 
in the high/supersoft state is $\sim$ a few $10^{39}$ 
erg s$^{-1}$, and that in the low/soft state 
is $\sim$ a few $10^{37}$ erg s$^{-1}$,
the long-term average luminosity may be 
$\sim 2 \times 10^{39}$.
This clearly raises two problems, related to the average 
long-term accretion rate and to the peak luminosity.
Since all the power must ultimately come from nuclear burning, 
the long-term average luminosity requires an average accretion rate 
$\sim (0.5$--$1) \times 10^{-5} M_{\odot}$ yr$^{-1}$.
From stellar evolution models \citep{rap05}, it appears that even late-type 
B stars (likely progenitor of the putative massive white dwarf 
and likely donor star in the system) can provide 
such extreme rates during thermal-timescale episodes of 
mass transfer, lasting $\sim 10^5$ yr.
If the same accretion rate was used to power an accreting BH, 
it would produce a bolometric luminosity  
$\sim (3$--$6) \times 10^{40}$ erg s$^{-1}$, 
which is similar to the maximum luminosity 
inferred for ULXs; the X-ray luminosity distribution 
of ULXs has a downturn at $L_{\rm X} \approx 2 \times 10^{40}$ 
erg s$^{-1}$ \citep{gri03,swa04}.
In other words, explaining the most luminous supersoft sources 
ever found as nuclear-burning sources with average long-term 
luminosity $\sim 10^{39}$ erg s$^{-1}$ requires mass-transfer 
conditions similar to those needed to explain the most luminous 
ULXs as accreting BHs with luminosities $\sim$ a few $10^{40}$ 
erg s$^{-1}$. We speculate that this may be more than 
a coincidence. Moreover, the alternative ULX scenario 
of a super-Eddington outflow from a stellar-mass BH (Section 4.2; 
\citet{car07,kin08}) also requires mass transfer rates 
of up to $\approx 10^{-5} M_{\odot}$ yr$^{-1}$ 
on the thermal timescale, possibly from a B-type donor star. 
Therefore, the required limits on the long-term average 
mass transfer rate are similar for both models.

As for the peak luminosity $\sim$ a few $10^{39}$ erg s$^{-1}$, 
this is highly super-Eddington for a white dwarf, and 
a factor of 10 above the luminosity produced by steady 
surface burning. However, steady burning cannot 
persist at accretion rates $\approx 10^{-5} M_{\odot}$ yr$^{-1}$.
We speculate that the high/supersoft phases may be transient 
super-Eddington events (fireball scenario), 
after which the photosphere shrinks again 
to the white-dwarf surface, and accretion resumes.
It was estimated \citep{sta04} 
that the hydrogen layer involved 
in surface nuclear burning has a mass 
$\sim 10^{-6} M_{\odot}$. Therefore, even if further hydrogen accretion 
is shut off (i.e., during the transient super-Eddington 
outburst), simply the complete burning of this layer 
can provide a luminosity $\ga 10^{39}$ erg s$^{-1}$ for 
several weeks, or $> 10^{40}$ erg s$^{-1}$ 
for several days, and the layer itself can be replenished 
in $\sim 1$ month, before another outburst. 
A similar process may be driving the outburst 
cycles of the supersoft ULX in M101 \citep{kon05}, 
which shows varying temperatures between 
$\approx 50$ and $\approx 150$ eV corresponding 
perhaps to phases of photospheric expansion 
and contraction. Super-Eddington outbursts 
powered by non-steady episodes of nuclear burning 
have been observed in some Novae, most notably LMC 91, 
which peaked at $\approx 2.6 \times 10^{39}$ erg s$^{-1}$ 
\citep{sch01}.

In conclusion, we suggest that a fireball white-dwarf model 
is still a viable scenario for this extreme source (and perhaps also 
for the whole class of supersoft ULXs), considering 
that its true bolometric luminosity is likely to be 
an order of magnitude less than originally estimated 
\citep{car07}, and that the required accretion rate 
$\approx 10^{-5} M_{\odot}$ yr$^{-1}$ is similar to the rate 
invoked for super-Eddington outflows in the most luminous ULXs.

\subsection{Time variability}

From a timing study of the {\it XMM-Newton} observation, 
it was found \citep{car07} that the X-ray emission 
has a modulation with an apparent period of $\approx 4.1$ hr.
It is not clear what can produce this period 
or timescale. If it was the binary period, 
from the period-density relation \citep{war95} 
we infer an average density $\approx 6$ g cm$^{-3}$ 
inside the Roche lobe of the donor star (the mean 
solar density is $\approx 1.4$ g cm$^{-3}$). 
This is consistent with a main-sequence star 
with a mass $\approx 0.5 M_{\odot}$.
Other well-known supersoft sources such as CAL\,87 
are known to have a low-mass donor with a binary 
period of a few hours \citep{cal89}. However, 
for NGC\,4631 X1, such a low-mass donor is inconsistent 
with the required mass transfer rates in both the stellar-mass 
BH and white dwarf scenarios, and is also at odds 
with the young age of the stellar population 
in this starburst galaxy 
(i.e., the galaxy is more likely to contain bright 
high-mass X-ray binaries). A B-type donor star 
is consistent with a massive white-dwarf compact object, 
both from theoretical arguments \citep{ibe94} 
and observationally \citep{ber00}; 
the lifetime of the B-type progenitor of a $1.3 M_{\odot}$ 
white dwarf is $\approx 50-70$ Myr. 

If the $4$-hr X-ray modulation is not the binary period, 
then it could be due either to a disk precession in the BH scenario, 
or to the rotational period of the white dwarf, 
or to pulsations in the donor star which affect 
the rate of mass transfer. A study of these scenarios 
is beyond the scope of this work. We just point out 
that $4$ hrs is the characteristic pulsation period 
of the B-type $\beta$-Cephei stars \citep{sta05}. 
Pulsations of a $\beta$-Cep donor have been invoked 
in the past as a possible cause of X-ray periodicities 
in some accreting binaries \citep{ber00,fin92}.



\begin{table}
\begin{center}
\caption{Best-fitting parameters for the {\it Chandra}/ACIS 
and {\it XMM-Newton}/EPIC spectra of X2. 
Spectral model: {\tt wabs*wabs*diskbb}. 
Values in brackets were fixed. Errors are 90\%
confidence levels for 1 interesting parameter ($\Delta \chi^2 =
2.7$). }
\vspace{0.2cm}
\begin{tabular}{lcr}
\tableline\tableline\\
Parameter & {\it Chandra} Value & {\it XMM-Newton} Value \\
\tableline\\
$N_{\rm {H,Gal}}$\tablenotemark{a} & $(1.3 \times 10^{20})$ & $(1.3 \times 10^{20})$\\[6pt]
$N_{\rm {H}}$ & $28.3^{+3.6}_{-3.2} \times 10^{21}$ & $26.4^{+3.5}_{-3.2} \times 10^{21}$\\[6pt]
$kT_{\rm{dbb}}$ (keV) & $1.49^{+0.22}_{-0.18}$ & $1.26^{+0.11}_{-0.10}$\\[6pt]
$K_{\rm{dbb}}$\tablenotemark{b} & $5.8^{+4.5}_{-2.6}  \times 10^{-3}$
           & $8.5^{+4.5}_{-2.8}  \times 10^{-3}$ \\[6pt]
\tableline\\
$\chi^2$/dof & $0.65 (42.4/65)$ & $0.78 (42.1/54)$ \\[3pt]
\tableline\\
$f_{\rm 0.3-10}$\tablenotemark{c} &$3.0^{+0.1}_{-0.2} \times 10^{-13}$ 
         &$2.1^{+0.1}_{-0.2} \times 10^{-13}$\\[6pt]
$L_{\rm 0.3-10}$\tablenotemark{d} &$4.0^{+0.3}_{-0.3} \times 10^{39}$
        & $3.1^{+0.2}_{-0.2} \times 10^{39}$ \\[6pt]
\tableline\\
\end{tabular}
\vspace{-0.2cm}
\tablenotetext{a}{From \citet{K05}. Units of cm$^{-2}$.}
\tablenotetext{b}{$K_{\rm{dbb}} = \left[R_{\rm in}({\rm km})/d(10{\rm ~kpc})\right]^2 
           \times \cos \theta$ where $R_{\rm in}$ is the apparent inner-disk radius 
           and $\theta$ the viewing angle; $\theta=0$ is face-on.}
\tablenotetext{c}{Observed flux in the $0.3$--$10$ keV band; 
          units of erg cm$^{-2}$ s$^{-1}$.}
\tablenotetext{d}{Unabsorbed luminosity in the $0.3$--$10$ keV band; 
          units of erg s$^{-1}$.}\\
\end{center}
\end{table}

\section{Other luminous X-ray sources}

\subsection{X2} 
Apparently coincident with a young star cluster, 
this highly-absorbed ULX 
($N_{\rm H} \sim 2 \times 10^{22}$ cm$^{-2}$), 
was not detected in any {\it ROSAT} observation (not surprisingly). 
This source can be classified as a ``convex-spectrum'' ULX
(using the terminology of \citet{mak07}).
The spectral curvature can be formally modelled with 
a standard disk-blackbody spectrum (Figure 7 and Table 4), with 
a color temperature 
$kT_{\rm in} \approx 1.3$--$1.5$ keV (in the {\it XMM-Newton} 
and {\it Chandra} data, respectively). However, 
as is generally the case in this class of ULXs, 
such temperatures are too high for the estimated 
luminosities ($\approx 3$ and $4 \times 10^{39}$ 
erg s$^{-1}$, respectively). One possibility is that 
the emission comes from a slim disk \citep{wat01} 
rather than a standard disk; if so, the mass accretion rate may be 
an order of magnitude above the Eddington rate, 
while the luminosity may be $\sim L_{\rm Edd}$, 
and the BH mass $\sim 20$--$30 M_{\odot}$ 
(for comparison with other sources in a similar state, 
see Fig.~3 in \citet{mak07}).
Alternatively, the convex spectrum may be modelled 
equally well with a Comptonized component, arising 
from a warm ($kT \sim 3$--$4$ keV) corona 
(see a comparison between the two scenarios 
in \citet{sto06}). Other purely 
phenomenological models such as a broken power-law
(breaking at $\sim 4$ keV) also provide good fits.

\begin{figure}
\includegraphics[angle=-90,width=1\columnwidth]{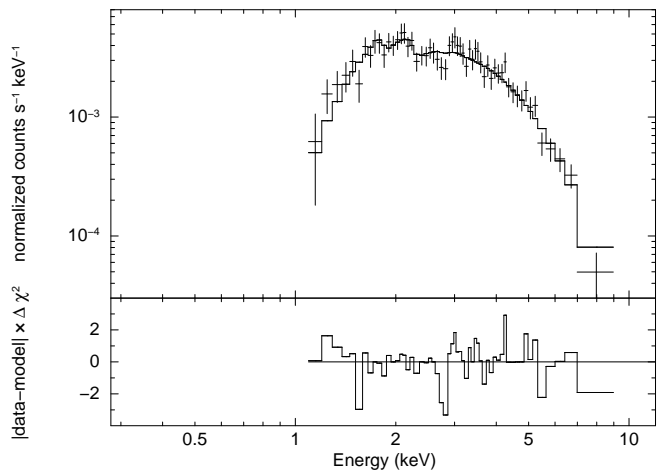}
\figcaption{{\it XMM-Newton}/EPIC spectrum of X2 (datapoints 
and $\chi^2$ residuals), 
fitted with a highly absorbed disk-blackbody model; see Table 4 
for the best-fitting parameters.}
\label{fig:figure7}
\vspace{0.3cm}
\end{figure}


\begin{figure}
\includegraphics[angle=-90,width=1\columnwidth]{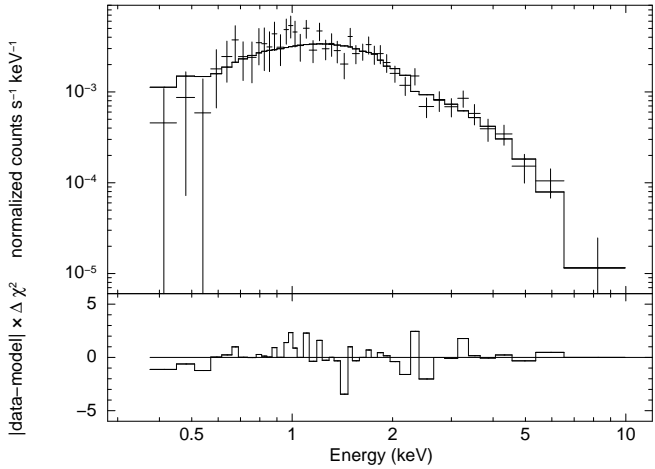}
\figcaption{{\it XMM-Newton}/EPIC spectrum of X3 (datapoints 
and $\chi^2$ residuals), 
fitted with a disk-blackbody model; see Table 5 
for the best-fitting parameters.}
\label{fig:figure8}
\vspace{0.3cm}
\end{figure}

\subsection{X3} 
This X-ray source is also well 
modelled with a disk-blackbody spectrum (Figure 8 and Table 5), with 
$kT{\rm in} \approx 1.2$--$1.4$ keV (in the {\it XMM-Newton} 
and {\it Chandra} data, respectively), similar 
to the parameters found for X2. However, its 
emitted luminosity in the $0.3$--$10$ keV band 
is only $\approx 4 \times 10^{38}$ erg s$^{-1}$, 
constant between {\it Chandra} and {\it XMM-Newton} 
and below the {\it ROSAT} detection limit. 
Therefore, X3 is consistent with  
a stellar-mass BH near the upper-luminosity end 
of its high/soft state. Assuming a spectral hardening 
factor $\approx 1.7$--$2$ \citep{shi95,gie04,sha06}, 
the standard-disk temperature-luminosity relation 
\citep{mak00} suggests a BH mass 
$\approx 5$--$7 M_{\odot}$; the fitted (apparent) 
inner-disk radius $R_{\rm in} \approx 30(\cos \theta)^{-0.5}$ 
km, as expected.

\subsection{X4} 
This ULX shows a state transition 
between the {\it Chandra} and {\it XMM-Newton} observations. 
In {\it Chandra}, we detect a faint, soft source 
($0.3$--$8$ keV count rate $= (1.44 \pm 0.16) \times 10^{-3}$ 
counts s$^{-1}$), 
well fitted ($\chi^2_{\nu} = 0.88$) by an optically-thin 
thermal-plasma model (Table 6) with a temperature 
$kT_{\rm rs} = 1.2^{+0.5}_{-0.2}$ keV.
Every other spectral model (power-law, disk-blackbody, 
or any Comptonization models) yields $\chi^2_{\nu} \ga 1.5$.
The emitted luminosity for the Raymond-Smith thermal-plasma 
model is $\approx 3 \times 10^{37}$ erg s$^{-1}$. 
In the {\it XMM-Newton} observations, there is a stronger, 
harder source at the same position, with a broad spectrum 
well fitted by a power-law plus disk-blackbody model (Figure 9 and Table 6). 
The power-law index is $\Gamma \approx 1.9$; there is 
no evidence of a steepening break or spectral curvature 
near or above $\sim 5$ keV, unlike what we noted for X2. 
The disk-blackbody component has a color 
temperature $kT_{\rm in} = 0.20^{+0.11}_{-0.05}$ keV.
The unabsorbed luminosity in the $0.3$--$10$ keV band 
is $(2.1 \pm 0.2) \times 10^{39}$ erg s$^{-1}$, 
$\approx 30$ per cent of which in the disk-blackbody component. 
It is possible that the thermal-plasma component 
seen in the {\it Chandra} dataset is also present 
in the {\it XMM-Newton} spectrum, although we cannot 
place strong constraints on it (Table 6).

In summary, X4 is a transient ``power-law'' ULX 
(using again the spectral classification of \citet{mak07}) 
with a thermal soft-excess at low energies.
A physical interpretation for this class of ULXs 
is that the standard 
optically-thick accretion disk is directly visible 
at large radii but is replaced or covered 
by a scattering-dominated region (producing a broader, 
power-law-like spectrum) at small radii 
(see, e.g., the review by \citet{sor08} 
and references therein). 
The apparent inner-disk radius (which 
we may identify with the transition radius between 
standard disk and Comptonizing regions) 
$R_{\rm in} \approx 1600(\cos \theta)^{-0.5}$ km.
From the relative contribution of thermal 
and (less radiatively efficient) power-law components, 
we speculate that this transition 
radius is $\ga$ a few times the innermost stable orbit; 
hence, the BH mass is likely to be $\la 50 M_{\odot}$.

Evidence of variability on monthly timescales for X4 
was already found in the series of {\it ROSAT} observations 
\citep{vog96}. We searched for short-term 
variability in the {\it XMM-Newton} observation (Figure 10), 
and found that a constant count rate is statistically ruled out 
($\chi^{2}_{\nu} =117.2/83$). However, we found no dominant 
frequency or other spectral features in its power-density spectrum.

\begin{table}
\begin{center}
\caption{Best-fitting parameters for the {\it Chandra}/ACIS 
and {\it XMM-Newton}/EPIC spectra of X3. 
Spectral model: {\tt wabs*wabs*diskbb}. 
Values in brackets were fixed. Errors are 90\%
confidence levels for 1 interesting parameter ($\Delta \chi^2 =
2.7$). }
\vspace{0.2cm}
\begin{tabular}{lcr}
\tableline\tableline\\
Parameter & {\it Chandra} Value & {\it XMM-Newton} Value \\
\tableline\\
$N_{\rm {H,Gal}}$\tablenotemark{a} & $(1.3 \times 10^{20})$ & $(1.3 \times 10^{20})$\\[6pt]
$N_{\rm {H}}$ & $2.0^{+1.0}_{-0.9} \times 10^{21}$ & $1.1^{+0.6}_{-0.5} \times 10^{21}$\\[6pt]
$kT_{\rm{dbb}}$ (keV) & $1.37^{+0.32}_{-0.21}$ & $1.20^{+0.17}_{-0.15}$\\[6pt]
$K_{\rm{dbb}}$\tablenotemark{b} & $7.8^{+6.9}_{-4.3}  \times 10^{-4}$
           & $13.8^{+9.2}_{-4.4}  \times 10^{-4}$ \\[6pt]
$\chi^2$/dof & $0.81 (17.9/22)$ & $0.75 (30.7/41)$ \\[3pt]
\tableline\\
$f_{\rm 0.3-10}$\tablenotemark{c} &$4.7^{+0.1}_{-0.1} \times 10^{-14}$ 
         &$5.0^{+0.1}_{-0.1}\times 10^{-14}$\\[6pt]
$L_{\rm 0.3-10}$\tablenotemark{d} &$3.9^{+0.2}_{-0.2} \times 10^{38}$
        & $4.0^{+0.2}_{-0.2} \times 10^{38}$ \\[6pt]
\tableline\\
\end{tabular}
\vspace{-0.2cm}
\tablenotetext{a}{From \citet{K05}. Units of cm$^{-2}$.}
\tablenotetext{b}{$K_{\rm{dbb}} = \left[R_{\rm in}({\rm km})/d(10{\rm ~kpc})\right]^2 
           \times \cos \theta$ where $R_{\rm in}$ is the apparent inner-disk radius 
           and $\theta$ the viewing angle; $\theta=0$ is face-on.}
\tablenotetext{c}{Observed flux in the $0.3$--$10$ keV band; 
          units of erg cm$^{-2}$ s$^{-1}$.}
\tablenotetext{d}{Unabsorbed luminosity in the $0.3$--$10$ keV band; 
          units of erg s$^{-1}$.}\\
\end{center}
\end{table}


\begin{table}
\begin{center}
\caption{Best-fit parameters for the {\it Chandra}/ACIS 
and {\it XMM-Newton}/EPIC spectra of X4. 
Spectral model: {\tt wabs*[(wabs$_{1}$*raymond-smith) 
+ wabs$_{2}$*(power-law+diskbb)]}. 
Values in brackets were fixed. Errors are 90\%
confidence levels for 1 interesting parameter ($\Delta \chi^2 = 2.7$). }
\vspace{0.2cm}
\begin{tabular}{lcr}
\tableline\tableline\\
Parameter  & {\it Chandra} Value & {\it XMM-Newton} Value \\
\tableline\\
$N_{\rm {H,Gal}}$\tablenotemark{a} & $(1.3 \times 10^{20})$ 
        & $(1.3 \times 10^{20})$ \\[6pt]
$N_{\rm {H, 1}}$ &$3.2^{+10.2}_{-3.2} \times 10^{20}$ 
        &  $(3.2 \times 10^{20})$ \\[6pt]
$N_{\rm {H, 2}}$ & -  
        &  $3.6^{+0.6}_{-0.6} \times 10^{21}$ \\[6pt]
$kT_{\rm{rs}}$ (keV) & $1.24^{+0.53}_{-0.18}$ 
        & $(1.24)$ \\[6pt]
$Z(Z_{\odot})$ & $(1.0)$ & $(1.0)$ \\[6pt]
$K_{\rm{rs}}$\tablenotemark{b} & $2.7^{+1.3}_{-1.0} \times 10^{-6}$ 
        & $2.2^{+4.3}_{-2.2} \times 10^{-6}$\\[6pt]
$\Gamma$\tablenotemark{c}  & - & $1.88^{+0.13}_{-0.12}$ \\[6pt]
$N_{\rm {pl}}$\tablenotemark{d} & - 
        &  $3.5^{+0.4}_{-0.3} \times 10^{-5}$ \\[6pt]
$kT_{\rm{dbb}}$ (keV) & - & $0.20^{+0.11}_{-0.05}$\\[6pt]
$K_{\rm{dbb}}$\tablenotemark{e} & - 
        & $4.5^{+2.7}_{-4.4}$ \\[6pt]
\tableline\\
$\chi^2$/dof & $0.88 (4.4/5)$ & $0.98 (67.7/69)$\\[3pt]
\tableline\\
$f_{\rm 0.3-10}$\tablenotemark{f} & $0.4^{+0.2}_{-0.2} \times 10^{-14}$ 
         &$15.4^{+1.0}_{-1.0} \times 10^{-14}$ \\[6pt]
$L_{\rm 0.3-10}$\tablenotemark{g} & $0.03^{+0.01}_{-0.01} \times 10^{39}$ 
         &$2.1^{+0.2}_{-0.2} \times 10^{39}$ \\[6pt]
\tableline\\
\end{tabular}
\vspace{-0.2cm}
\tablenotetext{a}{From \citet{K05}. Units of cm$^{-2}$.}
\tablenotetext{b}{Raymond-Smith model normalization $K_{\rm rs} =
10^{-14}/\{4\pi\,[d_A\,(1+z)]^2\}\,
\int n_e n_H {\rm{d}}V$, where $d_A$ is
          the angular size distance to the source (cm), $n_e$ is the electron
          density (cm$^{-3}$), and $n_H$ is the hydrogen density (cm$^{-3}$).}
\tablenotetext{c}{Photon index.}
\tablenotetext{d}{Units of photons keV$^{-1}$ cm$^{-2}$ s$^{-1}$, at 1 keV.}
\tablenotetext{e}{$K_{\rm{dbb}} = \left[R_{\rm in}({\rm km})/d(10{\rm ~kpc})\right]^2 
           \times \cos \theta$ where $R_{\rm in}$ is the apparent inner-disk radius 
           and $\theta$ the viewing angle; $\theta=0$ is face-on.}
\tablenotetext{f}{Observed flux in the $0.3$--$10$ keV band; 
          units of erg cm$^{-2}$ s$^{-1}$.}
\tablenotetext{g}{Unabsorbed luminosity in the $0.3$--$10$ keV band; 
          units of erg s$^{-1}$.}\\
\end{center}
\end{table}


\begin{table}
\begin{center}
\caption{Best-fitting parameters for the {\it Chandra}/ACIS 
and {\it XMM-Newton}/EPIC spectra of X5. 
Spectral model: {\tt wabs*wabs*power-law}. 
Values in brackets were fixed. Errors are 90\%
confidence levels for 1 interesting parameter ($\Delta \chi^2 =
2.7$). }
\vspace{0.2cm}
\begin{tabular}{lcr}
\tableline\tableline\\
Parameter & {\it Chandra} Value & {\it XMM-Newton} Value \\
\tableline\\
$N_{\rm {H,Gal}}$\tablenotemark{a} & $(1.3 \times 10^{20})$ & $(1.3 \times 10^{20})$\\[6pt]
$N_{\rm {H}}$ & $2.0^{+0.2}_{-0.2} \times 10^{21}$ & $2.5^{+0.2}_{-0.2} \times 10^{21}$\\[6pt]
$\Gamma$\tablenotemark{b}  & $1.79^{+0.08}_{-0.08}$ & $2.12^{+0.06}_{-0.03}$\\[6pt]
$N_{\rm {pl}}$\tablenotemark{c} &  $8.6^{+0.7}_{-0.7} \times 10^{-5}$ 
                  &  $13.7^{+0.8}_{-0.7} \times 10^{-5}$\\[6pt]
\tableline\\
$\chi^2$/dof & $0.95 (140.0/147)$ & $1.13 (187.1/166)$ \\[3pt]
\tableline\\
$f_{\rm 0.3-10}$\tablenotemark{d} &$4.2^{+0.2}_{-0.3} \times 10^{-13}$ 
       &$4.4^{+0.3}_{-0.2} \times 10^{-13}$\\[6pt]
$L_{\rm 0.3-10}$\tablenotemark{e} &$3.8^{+0.1}_{-0.1} \times 10^{39}$
        & $5.0^{+0.2}_{-0.2} \times 10^{39}$ \\[6pt]
\tableline\\
\end{tabular}
\vspace{-0.2cm}
\tablenotetext{a}{From \citet{K05}. Units of cm$^{-2}$.}
\tablenotetext{b}{Photon index.}
\tablenotetext{c}{Units of photons keV$^{-1}$ cm$^{-2}$ s$^{-1}$, at 1 keV.}
\tablenotetext{d}{Observed flux in the $0.3$--$10$ keV band; 
          units of erg cm$^{-2}$ s$^{-1}$.}
\tablenotetext{e}{Unabsorbed luminosity in the $0.3$--$10$ keV band; 
          units of erg s$^{-1}$.}\\
\end{center}
\end{table}

\subsection{X5} 
This ULX was in a luminous state 
in five of the six {\it ROSAT} observations 
\citep{vog96}. We found it again in a bright 
state during the {\it Chandra} and
{\it XMM-Newton} observations (Figure 11 and Table 7).
In both datasets, the X-ray spectra
are well fitted by a simple power-law of photon index
$\Gamma \sim 2$.
There is no statistically-significant evidence
of either a soft thermal component or a high-energy break
or downward curvature in the power-law. 
The unabsorbed isotropic luminosity in the $0.3$--$10.0$ keV band
is $(3.8 \pm 0.1) \times 10^{39}$ ~ergs~s$^{-1}$ ({\it Chandra})
and $(5.0 \pm 0.2) \times 10^{39}$ ~ergs~s$^{-1}$ {\it XMM-Newton}, 
only slightly higher than the range of estimated luminosities 
during the {\it ROSAT} observations \citep{liu05}, when the same 
{\it Chandra} or {\it XMM-Newton} model is applied 
to the {\it ROSAT} data\footnote{The {\it ROSAT}/HRI luminosity 
estimated by \citet{liu05} is a factor of 2 lower than 
our estimate because they assumed only line-of-sight 
absorption.}.
We did not find any variability within
the individual {\it Chandra} and {\it XMM-Newton} exposures.
In summary, X5 appears to be a typical power-law ULX, perhaps powered 
by accretion onto a BH with a mass $\la 50 M_{\odot}$.

\begin{figure}
\includegraphics[angle=-90,width=1\columnwidth]{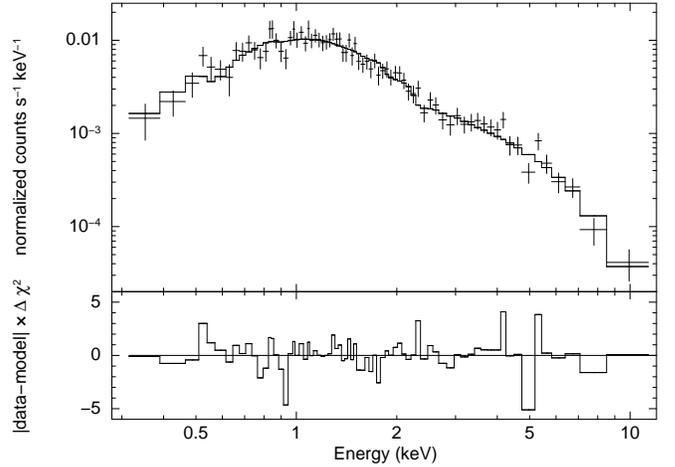}
\figcaption{{\it XMM-Newton}/EPIC spectrum of X4 (datapoints 
and $\chi^2$ residuals), 
fitted with a disk-blackbody plus power-law model; see Table 6 
for the best-fitting parameters.}
\label{fig:figure9}
\vspace{0.3cm}
\end{figure}


\begin{figure}
\includegraphics[angle=-90,width=1\columnwidth]{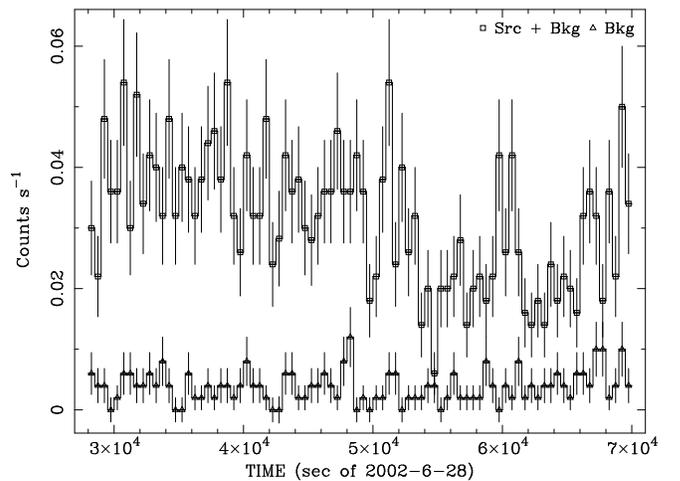}
\figcaption{{\it XMM-Newton}/EPIC lightcurve of X4, 
and corresponding background count rate, showing 
significant (aperiodic) short-term variability.}
\label{fig:figure10}
\vspace{0.3cm}
\end{figure}


\begin{figure}
\includegraphics[angle=-90,width=1\columnwidth]{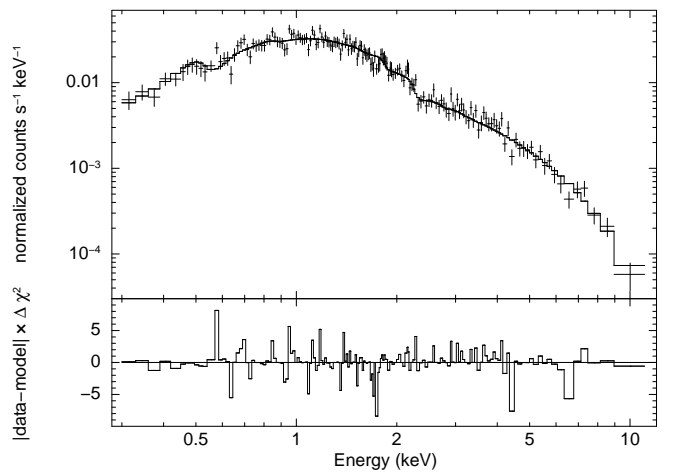}
\figcaption{{\it XMM-Newton}/EPIC spectrum of X5 (datapoints 
and $\chi^2$ residuals), 
fitted with a simple power-law model; see Table 7 
for the best-fitting parameters.}
\label{fig:figure11}
\vspace{0.3cm}
\end{figure}

\section{Conclusions}

We have studied the nature of the five brightest sources 
in NGC\,4631, using {\it XMM-Newton} and {\it Chandra} data.
Four of them can be classified as ULXs.
The most peculiar ULX, which we label X1, was previously studied 
by \citet{car07} and identified 
as a variable supersoft source with an apparent bolometric 
luminosity $\approx 3 \times 10^{40}$ erg s$^{-1}$. 
We re-examined the spectral data and found 
that in fact, its most likely luminosity may be 
only $\approx 4 \times 10^{39}$ erg s$^{-1}$; in fact, it could 
be even lower, if we consider that a blackbody approximation 
tends to overestimate the luminosity of supersoft sources.
This reduces the need for an intermediate-mass BH 
or other exotic scenarios. 

We found that when the source is in a low state 
({\it Chandra} observation), it appears soft (but 
not supersoft), consistent with a thermal spectrum 
at a temperature $\sim 0.1$--$0.3$ keV, and 
a luminosity $10^{37} \la L_{\rm bol} \la 10^{38}$ 
erg s$^{-1}$. We argued that this is inconsistent 
with an intermediate-mass BH. 
It is also unsual for a stellar-mass BH, which 
at those luminosities is expected to be in a power-law dominated 
low/hard state, or in a disk-dominated high/soft state with 
$T_{\rm in} \sim 0.5$--$1$ keV, 
based on our current knowledge of canonical accretion states.
As an alternative, we suggest that transient super-Eddington 
outbursts (fireball or photospheric expansion) powered 
by non-steady nuclear burning on the surface of a massive white dwarf
could be a viable scenario, as an extreme subclass of supersoft sources.
Outbursts due to photospheric expansion are expected 
when the accretion rate exceeds $\approx 10^{-6} M_{\odot}$ yr$^{-1}$.
Based on the sequence of available observations from 1991--2002,
the long-term average luminosity of the system
is $\approx (1$--$2) \times 10^{39}$ erg s$^{-1}$, which 
requires average accretion rates 
$\approx (5$--$10) \times 10^{-6} M_{\odot}$ yr$^{-1}$.
Although very high, such rates are achievable during phases 
of thermal-timescale mass transfer in B stars, 
and are similar to the rates required to explain 
the most luminous ULXs powered by BH accretion.
Hence, we speculate that transient outbursts in nuclear-burning, 
massive white dwarfs may also explain the few other 
supersoft ULXs (all highly variable) found in nearby galaxies. 
Some of those sources would be seen as quasi-soft sources 
(in the definition of Di Stefano \& Kong 2004) when they are 
in a low state. 

The origin of the $4$-hr X-ray variability remains 
unexplained, whatever the nature of the compact object. 
Given the high mass transfer rate, we would expect 
a B-type donor star filling its Roche lobe; 
however, a $4$-hr period does not allow for massive donors.
Alternatively, the variability may be due to an accretion disk 
precession, or the rotation of the white dwarf, 
or $\beta$-Cephei pulsations in the donor star.

The other four brightest sources in NGC\,4631 
are almost certainly {\it bone fide} BHs, in different accretion states. 
X2 ($L_{\rm X} \approx 3 \times 10^{39}$ erg s$^{-1}$ 
in the {\it XMM-Newton} observation)
is a highly absorbed ``convex-spectrum'' ULX; its X-ray 
spectrum may be interpreted as emission from a slim disk, 
or from a low-temperature (a few keV) Comptonizing region.
X3 ($L_{\rm X} \approx 4 \times 10^{38}$ erg s$^{-1}$) 
is a stellar-mass BH in its classical disk-dominated 
high/soft state. X4 ($L_{\rm X} \approx 3 \times 10^{39}$ 
erg s$^{-1}$) is a transient power-law ULX with a soft-excess 
at $kT \approx 0.20$ keV; for this class of objects, 
we may be seeing a standard disk 
outside a transition radius, completely replaced 
or covered by a Comptonizing region at smaller radii.
X5 ($L_{\rm X} \approx 5 \times 10^{39}$ erg s$^{-1}$)
is a pure power-law ULX, with no evidence for low-energy 
soft excess or high-energy steepening.
All of them are consistent with accreting stellar-mass 
BHs, with masses $\la 50 M_{\odot}$.
However, the relation between the different 
phenomenological states (high/soft, convex-spectrum, 
power-law, power-law with soft excess), 
and in particular whether those states 
are uniquely a function of the normalized accretion rate, 
remains a topic for further theoretical and observational 
investigations.

\begin{acknowledgements}
We thank the referee for a thoughtful review and interesting suggestions 
which have considerably improved the paper.
This research has made use of the NASA/IPAC Extragalactic Database (NED) which
 is operated by the Jet Propulsion Laboratory, California Institute of
 Technology, under contract with NASA;
and from the Chandra Data Archive, part of the Chandra X-Ray Observatory  
 Science Center (CXC) which is operated for NASA by SAO. Support for this research 
was provided in part by NASA under Grant NNG04GC86G issued through the Office of Space 
Science and from the Space Telescope Science Institute under the grant HST/AR-10954.
RS acknowledges support from a Leverhulme Fellowship, a UK-China Fellowship for excellence, and from Tsinghua University (Beijing). We thank Kinwah Wu, Shaung-Nan Zhang, Xin-Lin Zhou for their comments and discussions. 
\end{acknowledgements}













\begin{thebibliography}{}
\bibitem[Arnaud(1996)]{arn96} Arnaud, K. A. 1996, Astronomical Data 
     Analysis Software and Systems V, ASP Conference Series Vol.~101, 
     G. H. Jacoby and J. Barnes eds, 17
\bibitem[Begelman(2002)]{beg02} Begelman, M. C. 2002, ApJ, 568, L97
\bibitem[Begelman(2006)]{beg06}  Begelman, M. C. 2006, ApJ, 643, 1065
\bibitem[Bergh{\"o}fer et al.(2000)]{ber00} Bergh{\"o}fer, 
       T.~W., Vennes, S., \& Dupuis, J.\ 2000, \apj, 538, 854 
\bibitem[Boller et al.(2003)]{bol03} Boller, Th., Tanaka, Y., Fabian, A., 
       Brandt, W. N., Gallo, L., Anabuki, N., Haba, Y., \& Vaughan S. 
       2003, MNRAS, 343, L89
\bibitem[Callanan et al.(1989)]{cal89} Callanan, P.~J., 
       Machin, G., Naylor, T., \& Charles, P.~A.\ 1989, \mnras, 241, 37 
\bibitem[Carpano et al.(2007)]{car07}  Carpano, S., Pollock, A. M. T., 
     King, A. R., Wilms, J., \& Ehle, M. 2007, A\&A, 471, L55
\bibitem[Carpano et al.(2006)]{car06}  Carpano, S., Wilms, J., 
     Schirmer, M., \& Kendziorra, E. 2006, A\&A, 458, 747
\bibitem[Cash(1979)]{cas79} Cash, W. 1979, ApJ, 228, 939
\bibitem[Di Stefano \& Kong(2004)]{dis04} Di Stefano, R., 
    \& Kong, A. K. H. 2004, ApJ, 609, 710
\bibitem[Fabbiano et al.(2003)]{fab03} Fabbiano, G., King, A. R., 
    Zezas, A., Ponman, T. J., Rots, A., \& Schweizer, F. 2003, 
    ApJ, 591, 843
\bibitem[Fabbiano, Kim \& Trinchieri(1992)]{fab92} Fabbiano, G., 
       Kim, D.-W., \& Trinchieri, G. 1992, ApJS, 80, 531
\bibitem[Feng \& Kaaret(2005)]{fen05} Feng, H., 
       \& Kaaret, P. 2005, ApJ, 633, 1052
\bibitem[Finley et al.(1992)]{fin92} Finley, J.~P., Belloni, T., 
      \& Cassinelli, J.~P.\ 1992, \aap, 262, L25
\bibitem[Gierli{\'n}ski \& Done(2004)]{gie04} Gierli{\'n}ski, 
      M., \& Done, C.\ 2004, \mnras, 347, 885
\bibitem[Grimm, Gilfanov \& Sunyaev(2003)]{gri03} Grimm, H.-J., 
	Gilfanov, M., \& Sunyaev, R. 2003, MNRAS, 339, 793
\bibitem[Iben \& Tutukov(1994)]{ibe94} Iben, I.~J., 
     \& Tutukov, A.~V.\ 1994, \apj, 431, 264 
\bibitem[Kalberla et al.(2005)]{K05} Kalberla, P. M. W., Burton, W. B., 
     Hartmann, D., Arnal, E. M., Bajaja, E., Morras, R., 
     \& Poppel, W. G. L. 2005, A\&A, 440, 775
\bibitem[Kahabka(2004)]{kah04} Kahabka, P.\ 2004, \aap, 416, 57
\bibitem[Kahabka \& van den Heuvel(1997)]{kah97} Kahabka, P., 
     \& van den Heuvel, E. P. J. 1997, ARA\&A, 35, 69
\bibitem[Kennicutt(1998)]{ken98} Kennicutt, R. C. 1998, ARA\&A, 36, 189
\bibitem[King(2008)]{kin08} King, A. R. 2008, MNRAS, 385, L113
\bibitem[King et al.(2001)]{kin01} King, A. R., Davies, M. B., 
     Ward, M. J., Fabbiano, G., \& Elvis, M. 2001, ApJ, 552, L109
\bibitem[Kong \& Di Stefano(2003)]{kon03} Kong, A. K. H., 
    \& Di Stefano, R. 2003, ApJ, 590, L13
\bibitem[Kong \& Di Stefano(2005)]{kon05} Kong, A. K. H., 
    \& Di Stefano, R. 2005, ApJ, 632, 107L
\bibitem[Liu \& Bregman(2005)]{liu05} Liu, J.-F., \& 
    Bregman, J. N. 2005, ApJSS, 157, 59
\bibitem[Makishima(2007)]{mak07} Makishima, K.\ 2007, IAU Symposium, 
    238, 209
\bibitem[Makishima et al.(2000)]{mak00} Makishima, K., 
    et al. 2000, ApJ, 535, 632
 
\bibitem[Miller \& Colbert(2004)]{mil04} Miller, M. C., 
       \& Colbert, E. J. M. 2004, IJMPD, 13, 1
\bibitem[Ohsuga \& Mineshige(2007)]{Ohs07} Ohsuga, K., \& Mineshige, S. 2007, 
    ApJ, 670, 1283
\bibitem[Page et al.(2003)]{pag03} Page, M. J., Davis, S. W., 
         \& Salvi, N. J. 2003, MNRAS, 343, 1241
\bibitem[Pakull \& Mirioni(2002)]{pak02} Pakull, M. W., \& Mirioni, L. 2002, 
         in the unpublished proceedings of the symposium 
         ``New visions of the X-ray Universe in the {\it XMM-Newton} 
         and {\it Chandra} era'', ESTEC, 
          The Netherlands (November 2001), astro-ph/0202488
\bibitem[Poutanen et al.(2007)]{pou07} Poutanen, J., 
        Lipunova, G., Fabrika, S., Butkevich, A.~G., 
        \& Abolmasov, P.\ 2007, \mnras, 377, 1187
\bibitem[Rappaport, Podsiadlowski \& Pfahl(2005)]{rap05} Rappaport, S. A., 
       Podsiadlowski, Ph., \& Pfahl, E. 2005, MNRAS, 356, 401
\bibitem[Read, Ponman \& Strickland(1997)]{rea97} Read, A. M., 
       Ponman, T. J., \& Strickland, D. K. 1997, MNRAS, 286, 626
\bibitem[Schwarz et al.(2001)]{sch01} Schwarz, G. J., Shore, S. N., Starrfield, S., 
        Hauschildt, P. H., Della Valle, M., \& Baron, E. 2001, MNRAS, 320, 103
\bibitem[Seth, Dalcanton \& de Jong(2005)]{set05} Seth, A. C., 
      Dalcanton, J. J., \& de Jong, R. S. 2005, AJ, 129, 1331
\bibitem[Shafee et al.(2006)]{sha06} Shafee, R., McClintock, 
        J.~E., Narayan, R., Davis, S.~W., Li, L.-X., 
        \& Remillard, R.~A.\ 2006, \apjl, 636, L113 
\bibitem[Shakura \& Sunyaev(1973)]{sha73} Shakura, N. I., 
      \& Sunyaev, R. A. 1973, A\&A, 24, 337
\bibitem[Shimura \& Takahara(1995)]{shi95} Shimura, T., 
      \& Takahara, F.\ 1995, \apj, 445, 780
\bibitem[Soifer et al.(1989)]{soi89} Soifer, B. T., Boehmer, L., 
      Neugebauer, G., \& Sanders, D. B. 1989, AJ, 98, 766
\bibitem[Soria \& Kuncic(2008)]{sor08} Soria, R., \& Kuncic, Z. 2008, 
    to appear in the proceedings of the conference ``Observational evidence 
    of black holes'', Kolkata, India (February 2008), arxiv:0807.0016
\bibitem[Stankov \& Handler(2005)]{sta05} Stankov, A., \& Handler, 
      G.\ 2005, \apjs, 158, 193
\bibitem[Starrfield et al.(2004)]{sta04} Starrfield, S., Timmes, F. X., 
    Hix, W. R., Sion, E. M., Sparks, W. M., \& Dwyer, S. J., 2004, 
    ApJ, 612, L53
\bibitem[Stobbart et al.(2006)]{sto06} Stobbart, A.-M., 
       Roberts, T.~P., \& Wilms, J.\ 2006, \mnras, 368, 397 
\bibitem[Strickland et al.(2004a)]{str04a} Strickland, D. K., 
    Heckman, T. M., Colbert, E. J. M., Hoopes, C. G., \&
    Weaver, K. A. 2004b, ApJS, 151, 193
\bibitem[Strickland et al.(2004b)]{str04b} Strickland, D. K., 
    Heckman, T. M., Colbert, E. J. M., Hoopes, C. G., \&
    Weaver, K. A. 2004b, ApJ, 606, 829 
\bibitem[Swartz et al.(2002)]{swa02} Swartz, D. A., Ghosh, K. K., 
     Suleimanov, V., Tennant, A. F., \& Wu, K. 2002, ApJ, 574, 382 
\bibitem[Swartz et al.(2004)]{swa04} Swartz, D. A., Ghosh, K. K., 
	Tennant, A. F., \& Wu, K. 2004, ApJS, 154, 519
\bibitem[Tennant(2006)]{ten06} Tennant, A. F. 2006, AJ, 132, 1372
\bibitem[T\"ullmann et al.(2006a)]{tul06a} T\"ullmann, R., 
    Pietsch, W., Rossa, J., Breitschwerdt, D., \& Dettmar, R. J. 2006, 
    A\&A, 448, 43
\bibitem[T\"ullmann et al.(2006b)]{tul06b} T\"ullmann, R., 
    Breitschwerdt, D., Rossa, J., Pietsch, W., \& Dettmar, R. J. 2006, 
    A\&A, 457, 779
\bibitem[Vogler \& Pietsch(1996)]{vog96} Vogler, A., 
      \& Pietsch, W. 1996, A\&A, 311, 35
\bibitem[Wang et al.(2001)]{wan01} Wang, Q. D., Immler, S., 
     Walterbos, R., Lauroesch, J. T., \& Breitschwerdt, D. 2001, 
     ApJ, 555, L39
\bibitem[Warner(1995)]{war95} Warner, B.\ 1995, \apss, 232, 89 
\bibitem[Watarai, Mizuno \& Mineshige(2001)]{wat01} Watarai, K.-Y., 
         Mizuno, T., \& Mineshige, S.\ 2001, \apjl, 549, L77 
\bibitem[Winter, Mushotzky \& Reynolds(2006)]{win06} Winter, L. M., 
         Mushotzky, R. F., \& Reynolds, C. S. 2006, 649, 730
\bibitem[Winter, Mushotzky \& Reynolds(2007)]{win07} Winter, L. M., 
         Mushotzky, R. F., \& Reynolds, C. S. 2007, ApJ, 655, 163
\end{thebibliography}
\end{document}